\documentclass[11pt]{article}
\usepackage[margin=1in]{geometry}
\usepackage{authblk}
\usepackage[numbers]{natbib}
\usepackage{hyperref}

  \providecommand{\jnlcitation}[1]{}
  \providecommand{\authormark}[1]{}
  \providecommand{\titlemark}[1]{}
  \providecommand{\address}[2]{}
  \providecommand{\corres}[1]{}
  \providecommand{\presentaddress}[1]{}
  \providecommand{\keywords}[1]{}

\usepackage{enumitem}
\usepackage[utf8]{inputenc}
\usepackage{amsmath, amssymb}
\usepackage{graphicx}
\usepackage{subcaption}
\usepackage{soul}

\usepackage{booktabs}
\usepackage{xcolor}

\newcommand{\bZ}{\boldsymbol{Z}}
\newcommand{\btheta}{\boldsymbol{\theta}}

\usepackage{algorithm}
\usepackage{algpseudocode} 
\usepackage{amsthm}
\newtheorem{assumption}{Assumption}
\usepackage{etoolbox}
\AtBeginEnvironment{tabular}{\small}
\AtBeginEnvironment{tabular*}{\small}
\AtBeginEnvironment{tabularx}{\small}

\begin{document}
\title{Correcting Measurement Error and Zero Inflation in Functional Covariates for Scalar-on-Function Quantile Regression}
\author[1]{Caihong Qin}
\author[2]{Lan Xue}
\author[3]{Ufuk Beyaztas}
\author[1]{Roger S. Zoh}
\author[4]{Mark Benden}
\author[5]{Jeff Goldsmith}
\author[1]{Carmen D. Tekwe}

\affil[1]{Department of Epidemiology and Biostatistics, Indiana University School of Public Health}
\affil[2]{Department of Statistics, Oregon State University}
\affil[3]{Department of Statistics, Marmara University}
\affil[4]{Department of Environmental and Occupational Health, Texas A\&M University School of Public Health}
\affil[5]{Department of Biostatistics, Columbia University Mailman School of Public Health}


\date{}
\maketitle

\begin{abstract}
Wearable devices collect time-varying biobehavioral data, offering opportunities to investigate how behaviors influence health outcomes. However, these data often contain measurement error and excess zeros (due to nonwear, sedentary behavior, or connectivity issues), each characterized by subject-specific distributions. Current statistical methods fail to address these issues simultaneously. 
We introduce a novel modeling framework for zero-inflated and error-prone functional data by incorporating a subject-specific time-varying validity indicator that explicitly distinguishes structural zeros from intrinsic values. We iteratively estimate the latent functional covariates and zero-inflation probabilities via maximum likelihood, using basis expansions and linear mixed models to adjust for measurement error. To assess the effects of the recovered latent covariates, we apply joint quantile regression across multiple quantile levels. 
Through extensive simulations, we demonstrate that our approach significantly improves estimation accuracy over methods that only address measurement error, and joint estimation yields substantial improvements compared with fitting separate quantile regressions. 
Applied to a childhood obesity study, our approach effectively corrects for zero inflation and measurement error in step counts, yielding results that closely align with energy expenditure and supporting their use as a proxy for physical activity.
\end{abstract}    




\section{Introduction}\label{sec:intro}

\begin{figure}[!htbp]
    \centering
    \includegraphics[width=0.95\textwidth]{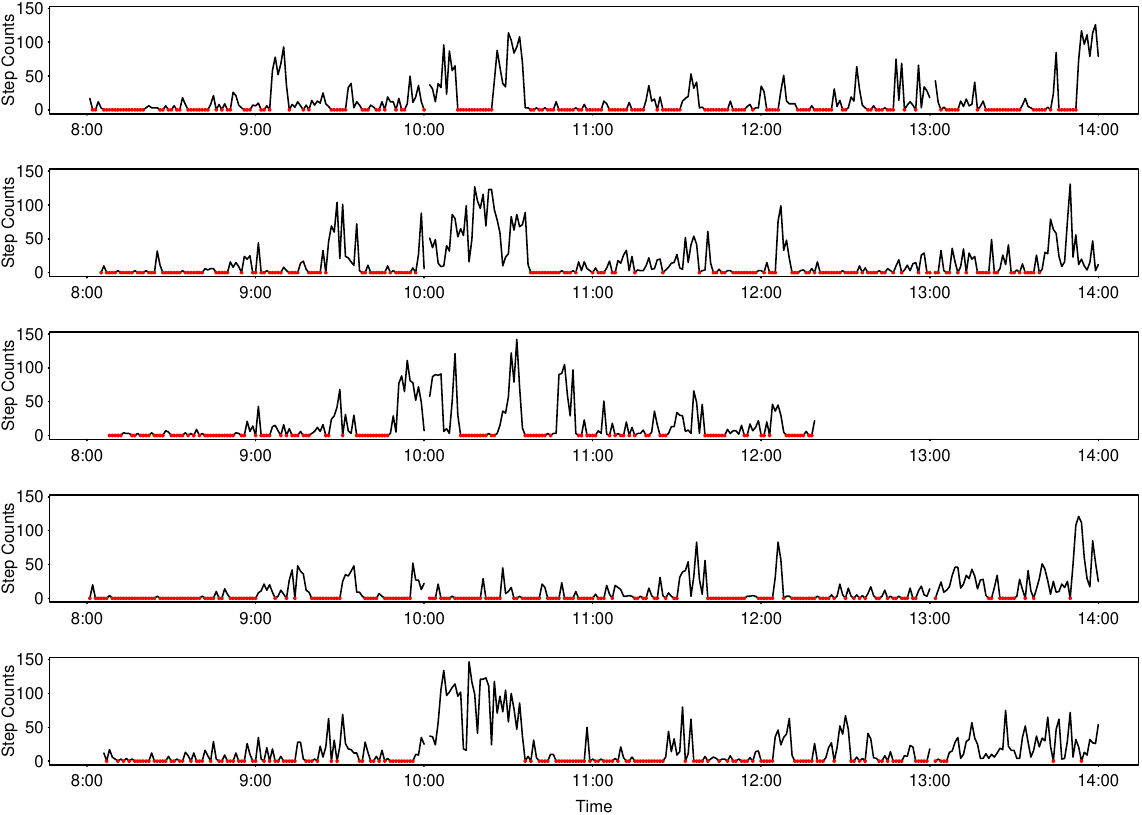}
    \caption{Minute-level step counts from one student over five different days from 8:00--14:00 in a childhood obesity study \citep{benden2014evaluation}. Red points represent zero step counts.}
    \label{fig:step_counts}
\end{figure}

In recent years, wearable technology has advanced rapidly, enabling researchers and clinicians to collect time-varying, high-dimensional biobehavioral data at an unprecedented scale. These data are increasingly prevalent across diverse fields, including epidemiology, behavioral science, and personalized health management \citep{seneviratne2017survey,banerjee2018wearable,vijayan2021review}. Despite the improved accuracy of device-based measurements compared to self-reported data, device-measured data still suffer from various sources of error \citep{feito2012effects,case2015accuracy,an2017valid,o2020well}, and numerous analytical challenges remain. 
Measurement error in wearable devices, which may arise from device limitations, variability in predicting intensity levels, and imperfect calibration equations \citep{bassett2012device,warolin2012effect,jadhav2022function}, can be heteroscedastic or correlated over time.
Moreover, wearable device data often exhibit excess zeros, i.e., zero inflation \citep{loeys2012analysis}, arising from periods of non-wear, sedentary behavior, or connectivity issues \citep{cho2021factors,cho2021identifying}, making interpretation difficult. These situations cannot be treated the same because zeros from sedentary behavior are not simply missing data; they provide valuable information on the user's physical activity \citep{mailey2014influence}.
For example, Figure~\ref{fig:step_counts} illustrates minute-level step counts collected over five weekdays (Monday–Friday, 8:00 AM to 2:00 PM) for one student in a classroom-based intervention study on childhood obesity \citep{benden2014evaluation}. These repeated daily measurements are treated as noisy replicates of an underlying latent activity curve and can be combined to estimate a single subject-specific latent step count trajectory. Periods without any plotted points (e.g., the afternoon on Day 3) indicate missing data, which are excluded from further analysis. The figure highlights the pronounced variability and the presence of zeros in physical activity data.
However, it is often challenging to distinguish the true cause of zero steps in real-world data. For instance, Jeffries
et al \citep{jeffries2014physical} found no significant difference in step counts between sedentary and non-wear bouts for some monitors, while both Vert et al \citep{vert2022detecting} and Randhawa et al \citep{randhawa2023statistical} observed that data recorded during non-wear periods can appear similar to data collected during sleep or sedentary periods.

Given these measurement complexities, understanding how these device-measured behaviors influence health outcomes is central to advancing public health research and guiding clinical recommendations. Previous studies have linked wearable-derived measures to outcomes such as obesity, type~2 diabetes, and cardiovascular health \citep{mcdonough2021health,rodriguez2021mobile,dhingra2023use,hughes2023wearable}, often treating these measures as functional data. 
However, many existing analyses overlook measurement error or zero inflation in the measures \citep{vijayan2021review,acar2024functional}.
Developing statistical methods that account for both measurement error and zero inflation is therefore essential for accurately quantifying associations between time-varying behaviors and health outcomes. 
While Wang et al.~\citep{wang2024generalized} proposed a truncation-based method, its threshold-based assumption for zeros is incompatible with physical activity data, where zeros may arise from either device non-wear or true inactivity, and it does not support the functional data structures commonly encountered in wearable studies.
To the best of our knowledge, limited methods currently exist in the statistical literature to simultaneously address these issues.

Various strategies have aimed to accommodate excess zeros in physical activity data. Bai et al \citep{bai2018two} modeled a time-dependent binary process to indicate whether a subject is active at a time interval, using thresholded observations. Li et al \citep{li2018three} expanded this approach to three categories (inactive, partially active, and active). However, as Randhawa et al \citep{randhawa2023statistical} noted, modern accelerometers continue recording even when users are not wearing them, resulting in the risk of misclassifying true sedentary periods as non-wear. Such misclassification can affect estimates of total physical activity, sedentary time, and associated health outcomes \citep{vanhelst2019comparison}. One common method of identifying non-wear time is to set a threshold for a certain number of consecutive zero counts \citep{janssen2015objective}. While this threshold can substantially influence estimates of sedentary time \citep{janssen2015objective,vanhelst2019comparison}, recommended values vary across studies, examples include 20 minutes \citep{esliger2005standardizing,janssen2015objective}, 30 minutes \citep{vanhelst2019comparison}, 45–60 minutes \citep{aadland2018comparison}, 60 minutes \citep{chinapaw2014sedentary}, or allowing for short intervals of nonzero counts \citep{troiano2008physical,choi2011validation}. Other approaches rely on additional data sources, such as activity diaries \citep{ottevaere2011use}, heart rate \citep{jeffries2014physical}, temperature \citep{vert2022detecting,skovgaard2023generalizability}, or raw three-dimensional accelerometry signals \citep{syed2020evaluating}. 
Given the lack of a unified standard for consecutive-zero thresholds, the frequent unavailability of supplementary data, and the presence of measurement error alongside zero inflation, it is crucial to develop new estimation methods that properly address both zero inflation and measurement error to accurately quantify physical activity levels.

In parallel, scalar-on-function regression has emerged as a powerful tool for linking functional data, such as time-varying physical activity measurements, to scalar outcomes. This model has been extensively studied, with comprehensive reviews provided in the literature \citep{ramsay2005,morris2015functional,wang2016functional,reiss2017methods}. 
However, classical functional data analysis often regards device-measured processes as smooth trajectories with random noise, which may be inadequate for covariates subject to measurement error and zero inflation.
Traditional measurement error corrections often rely on the assumption of independent or white-noise error structures \citep{yao2005functional,cardot2007smoothing,goldsmith2013corrected,reiss2017methods}, which do not reflect the correlated errors found in many wearable data sets. Some recent strategies have begun to address correlated measurement error through methods such as  multivariate joint regression calibration \citep{crainiceanu2009generalized,chakraborty2017regression} and simulation-extrapolation \citep{cai2015methods}; however, these methods assume that the error processes are shared across subjects and not designed to address zero inflation. Additionally, instrumental variable methods \citep{tekwe2019instrumental,tekwe2022estimation,zoh2024bayesian} and repeated measures approaches \citep{zhang2023partially,chen2024adjusting} have also been employed to address measurement error in functional data settings. Thus, existing methods  assume homoscedastic and uncorrelated errors, or treat measurement error as a Gaussian process, without accounting for zero inflation. Such assumptions may lead to biased estimation and incomplete insights into the relationships between complex behavioral processes and health endpoints.

Quantile regression \citep{koenker1978regression,koenker2005quantile} has gained increasing attention for analyzing health outcomes in observational studies, as it goes beyond mean responses to investigate how covariates affect different parts of the outcome distribution \citep{geraci2016qtools,beets2016physical,blankenberg2016troponin}. Cardot et al \citep{cardot2005quantile}, Kato \citep{kato2012estimation} and Yao et al \citep{yao2017regularized} have recently applied quantile regression to functional data. Wang et al \citep{wang2012corrected} and Chen and Müller \citep{chen2012conditional} studied quantile regression with covariates prone to measurement error. By modeling multiple quantiles jointly,  Wei and Carroll \citep{wei2009quantile} and Firpo et al \citep{firpo2017measurement} gain efficiency and ensure monotonic, non-crossing quantile functions, introducing additional complexities. Despite extensive methodological development in functional quantile regression, real-world applications of joint functional quantile regression models have largely overlooked bias correction for measurement error and zero inflation in the functional covariates.

In this paper, we propose a novel scalar-on-function quantile regression (SoFQR) framework to address the challenges posed by zero-inflated and error-prone functional covariates, as commonly encountered in wearable and biomedical data. Our approach introduces a subject-specific, time-varying validity indicator that explicitly distinguishes structural zeros (arising from device nonwear or technical issues) from genuine zeros due to inactivity. We jointly estimate the latent functional covariates and zero-inflation probabilities via an iterative maximum likelihood procedure, leveraging basis function expansions and linear mixed models to correct for measurement error. This contrasts with existing methods, which often assume homogeneous error structures and do not explicitly address zero inflation. To evaluate the impact of the recovered latent covariates across the outcome distribution, we employ a joint quantile regression approach that ensures coherence and stability across quantile levels. Together, these innovations enable our framework to flexibly accommodate heterogeneous, zero-inflated, and error-contaminated functional data, improving inference on functional predictors of scalar outcomes.

The remainder of this paper is organized as follows. In Section~\ref{sec:method}, we detail the proposed measurement error and zero-inflation correction procedure under a SoFQR framework. Section~\ref{sec:simu} presents comprehensive simulation studies to assess the estimation performance of our proposed methods and to quantify the advantages of joint quantile modeling. In Section~\ref{sec:real}, we apply our approach to a real-world childhood obesity dataset, highlighting its ability to yield actionable insights into the effects of time-varying behaviors on BMI across different quantile levels. Finally, Section~\ref{sec:conclusion} concludes with a summary of our findings and outlines potential directions for future research.

\section{Scalar-on-Function Quantile Regression}\label{sec:method}

We introduce a SoFQR framework \citep{cardot2005quantile,kato2012estimation} to link a scalar outcome \(Y\) to a functional covariate \(X(t)\). Let \(\{Y_i, X_i(t), \bZ_i\}_{i=1}^n\) be i.i.d.\ realizations of \((Y, X(t),\bZ)\), where \(X(t)\) is a centered square-integrable functional covariate on \(\mathcal{T}=[0,1]\) with \(\mathbb{E}[X(t)] = 0\), and \(\bZ \in \mathbb{R}^p\) is a vector of additional error-free covariates (including an intercept).
For a fixed quantile level \(\tau\in(0,1)\), the conditional \(\tau\)-th quantile of \(Y\) given \(X(t)\) and \(\bZ\) is 
$Q_{Y}(\tau \mid X, \bZ) = F_{Y}^{{-1}}(\tau \mid X, \bZ)$,
where $F_{Y}(\tau \mid X, \bZ)$ 
is the cumulative distribution function of $Y$ conditional on $X(t)$ and $\bZ$. In our SoFQR model, we assume
\begin{equation}\label{eq:fqm}
    Q_{Y_i}(\tau \mid X_i, \bZ_i)
    \;=\;
    \int_0^1 \beta\bigl(t,\tau\bigr)\,X_i(t)\,dt
    \;+\;
    \bZ_i^T\,\btheta(\tau),
\end{equation}
where \(\beta(\cdot,\tau)\in L_2([0,1])\) is the functional coefficient capturing how the trajectory of \(X(t)\) affects the \(\tau\)-th quantile of \(Y\), and \(\btheta(\tau)\in\mathbb{R}^p\) are scalar coefficients for the covariates in \(\bZ_i\). 

The SoFQR framework presented in model \eqref{eq:fqm} has received increasing attention in the functional data analysis literature. It was first formulated by Cardot et al \citep{cardot2005quantile} as a natural extension of classical linear quantile regression \citep{koenker1978regression}, who proposed a penalized spline estimator for the functional coefficient $\beta(t,\tau)$. Later, Kato \citep{kato2012estimation} developed an estimator based on functional principal component analysis {for SoFQR (without additional scalar covariates)} and established its optimal convergence rate in a minimax sense. Yao et al \citep{yao2017regularized} extended the model by considering a regularized partially functional quantile regression incorporating high-dimensional scalar covariates, with emphasis on variable selection. Similarly, Ma et al \citep{ma2019quantile} studied a high-dimensional model involving multiple functional covariates and introduced double penalization methods to select both functional and scalar predictors. In contrast to these regularization and basis expansion-based approaches, Ferraty et al \citep{ferraty2005conditional} and Chen and Müller \citep{chen2012conditional} proposed estimating the conditional quantile function by directly inverting the conditional distribution function, demonstrating the consistency properties of their estimators. More recently, the statistical inference procedures for model \eqref{eq:fqm} has been studied by Li et al \citep{li2022inference} and Sang et al \citep{sang2022statistical}. Specifically, Li et al \citep{li2022inference} proposed an adjusted Wald test for testing whether the regression parameter is constant across quantile levels, and derived its asymptotic chi-square distribution. Sang et al \citep{sang2022statistical} established the convergence rate and weak convergence of their estimator, and constructed both pointwise and simultaneous confidence intervals for the coefficient function.

This paper addresses a more challenging scenario in which the functional covariate 
\(X_i(t)\)  is latent and not directly observable. However, it may be measured by error prone proxy measures denoted by \(\{W_{ij}(t)\}_{j=1}^{J}\). Furthermore, the observed measures, $\{W_{ij}(t)\}_{j=1}^{J}$, may exhibit zero inflation. It is a characteristic commonly seen in digital health data or wearable device outputs, such as step counts, where periods of inactivity or device non-wear can lead to an overabundance of zeros. This phenomenon is also prevalent in certain biomedical markers, where physiological thresholds or detection limits result in frequent zero readings. Zero inflation can complicate statistical modeling and must be carefully accounted for to avoid biased inference.

\subsection{Model Assumptions on Repeated Measurements}\label{subsec:model-assumptions}

In this subsection, we introduce our modeling framework for repeated measurements subject to zero inflation and measurement error. We consider a setting in which each subject \(i\) has an underlying functional covariate \(X_i(t)\), and we collect \(J\) replicated measurements of this covariate at time \(t\). Denote these observed measurements by \(\{W_{ij}(t)\}_{j=1}^{J}\), recorded over different days or observational windows indexed by \(j = 1,\ldots,J\). 
{In practice, $W_{ij}(t)$ is recorded at sufficiently dense discrete time points so that it can be treated as functional data, as in our application with minute-level step count trajectories.}
We introduce a random validity indicator \(V_{ij}(t)\), specifying whether the measurement at time \(t\) is valid (effectively ``on'') or invalid due to device nonwear or connectivity issues. Let \(W_{ij}^*(t)\) represent the unobserved surrogate measurement recorded only when the device is valid.

To address the combined challenges of subject-specific heterogeneity in zero inflation and measurement error, we impose the following assumptions.

\begin{assumption}\label{assump:decomp}
We assume
\begin{equation}\label{equ:w}
    W_{ij}(t)
    \;=\;
    V_{ij}(t)\;W_{ij}^*(t),
\end{equation}
where 
\[
  V_{ij}(t)\sim\operatorname{Bernoulli}\bigl(1-\pi_i(t)\bigr),
\]
in which \(\pi_i(t)\), referred to as the zero-inflation probability, denotes the probability that the device produces a structural zero at time \(t\), due to non-wear or device issues.
\end{assumption}

\begin{assumption}\label{assump:expfam}
The unobserved surrogate \(W_{ij}^*(t)\) follows an exponential family distribution \(f\{w \mid X_i(t)\}\) with mean \(\mathbb{E}\bigl[W_{ij}^*(t)\mid X_i(t)\bigr] = X_i(t)\).
\end{assumption}

\begin{assumption}\label{assump:indep}
The validity indicator \(V_{ij}(t)\) is independent of the measurement error \(U_{ij}(t)\).
\end{assumption} 

\begin{assumption}\label{assump:piecewise}
The zero-inflation probability \(\pi_i(t)\) is subject-specific and time-varying, and is assumed to be piecewise constant over \([0,1]\). Specifically, the domain is divided into \(M\) segments,
\begin{equation}\label{eq:pw_pi}
    \pi_i(t)
    \;=\;
    \pi_{im}
    \quad
    \text{for }
    t \in \mathcal{T}_m = (t_{m-1},\,t_m],
    \quad
    m=1,\dots,M,
\end{equation}
with \(t_0 = 0\) and \(t_M = 1\).
\end{assumption}

Under Assumption~\ref{assump:decomp}, the probability of observing a zero measurement at time \(t\) is 
\[
    \mathbb{P}\bigl\{\,W_{ij}(t)=0\,\bigr\}
    \;=\;
    \underbrace{\mathbb{P}\bigl\{\,V_{ij}(t)=0\,\bigr\}}_{\text{structural zeros due to non-wear or device issues}} 
    \;+\; 
    \underbrace{\mathbb{P}\bigl\{\,W_{ij}^*(t)=0 \,\bigm|\; V_{ij}(t)=1\,\bigr\}\,\mathbb{P}\bigl\{\,V_{ij}(t)=1\,\bigr\}}_{\text{sampling zeros from the measurement distribution}}.
\]
When \(W_{ij}^*(t)\) is continuously distributed (e.g., Gaussian), the probability of an intrinsic zero is essentially zero, so zero measurements arise primarily from device-off times, simplifying the decomposition of observed zeros. However, if \(W_{ij}^*(t)\) follows a discrete distribution (e.g., Poisson), it may yield zeros even when the device is operational, making it difficult to distinguish between zeros arising from the measurement process and those due to structural causes such as non-wear. This introduces challenges in model estimation, forming a central focus of this paper.

Assumption~\ref{assump:expfam} highlights that \(W_{ij}^*(t)\) follows an exponential family distribution with mean \(X_i(t)\), thus linking the underlying functional covariate directly to the distribution of observed measurements when the device is valid. 
{We then define the measurement error as $U_{ij}(t) = W_{ij}^*(t) - X_i(t)$, so that $\mathbb{E}\bigl\{U_{ij}(t)\mid X_i(t)\bigr\} = 0$.}
This setup allows the covariance structure of $U_{ij}(t)$ to vary across subjects and depend on the true covariate $X_i(t)$, reflecting subject-specific device usage patterns and within-subject temporal correlation.
The uncorrelatedness assumed in Assumption~\ref{assump:indep} enables a factorization of the joint likelihood and is crucial for identifying and consistently estimating the latent functional covariate \(X_i(t)\).

The subject-specific, piecewise-constant \(\pi_i(t)\) in Assumption~\ref{assump:piecewise} is flexible and aligns with practical situations. Although the segmentation \(\{t_0,\ldots,t_M\}\) could also be subject-specific, for notational simplicity we use the same segmentation across all subjects. By segmenting the time axis, we gain flexibility in modeling device on/off patterns while controlling model complexity. This allows the zero-inflation probability to change at predetermined intervals, capturing shifts in factors such as device performance or user behavior over time. 
While the true zero-inflation process may vary smoothly, this piecewise constant approximation provides a tractable and interpretable framework that captures critical temporal variation in the data.

These assumptions collectively enable us to address zero inflation, measurement error, and subject-specific heterogeneity in a unified modeling framework. They are essential for ensuring that the subsequent estimation procedure remains valid and that the latent \(X_i(t)\) can be consistently recovered from the observed zero-inflated data.

\subsection{Estimation Procedure}\label{subsec:est-procedure}

We propose a two-stage estimation approach for recovering the latent \(X_i(t)\) and fitting a SoFQR model. In the first stage, we correct for zero inflation and measurement error to obtain an estimate of \(X_i(t)\), which yields \(\widehat{X}_i(t)\) based on the repeated measures \(\{W_{ij}(t)\}_{j=1}^J\). In the second stage, we plug the corrected \(\widehat{X}_i(t)\) into model \eqref{eq:fqm} to estimate regression coefficients \(\beta(\cdot,\tau)\) and \(\btheta(\tau)\).

In the first stage, our goal is to estimate both the zero-inflation probabilities \(\{\pi_i(t)\}\) and the latent curves \(\{X_i(t)\}\). Under an exponential-family model \(f\bigl(w\mid {X}_i(t)\bigr)\), the likelihood contribution of \(W_{ij}(t)\) at time \(t\) given \({X}_i(t)\) is
\[
    \ell_{ij}\bigl\{t \mid {X}_i(t)\bigr\}
    \;=\;
    \Bigl\{\,
        \pi_i(t)
        + \bigl(1-\pi_i(t)\bigl)\,f\bigl(0\mid {X}_i(t)\bigr)
    \Bigr\}^{\mathbb{I}\{W_{ij}(t)=0\}}
    \Bigl\{\,
        \bigl(1-\pi_i(t)\bigl)\,f\bigl(W_{ij}(t)\mid {X}_i(t)\bigr)
    \Bigr\}^{\mathbb{I}\{W_{ij}(t)>0\}},
\]
where \(\mathbb{I}\{\cdot\}\) denotes the indicator function. The zero-inflation probabilities \(\{\pi_i(t)\}\) can then be estimated by maximizing the likelihood of the observed data \(\{W_{ij}(t)\}_{j=1}^J\). Under the Assumption \ref{assump:piecewise} that \(\pi_i(t)\) is piecewise constant over \(\mathcal{T}_m\), we can obtain $\hat{\pi}_{im}$ by maximizing the likelihood for the interval $\mathcal{T}_m$, $\ell_i^{(m)}\bigl\{\pi_{im}\mid {X}_i(t)\bigl\}$, with
\[
    \ell_i^{(m)}\bigl\{\pi_{im}\mid {X}_i(t)\bigl\}
    \;=\;
    \prod_{t\in\mathcal{T}_m}\;\prod_{j=1}^J\;
    \ell_{ij}\bigl\{t \mid {X}_i(t),\,\pi_i(t)=\pi_{im}\bigr\}.
\]

Since \({X}_i(t)\) is latent, we propose to provide an initial estimation of \(\{{\pi}_{i}(t)\}\) to obtain a working estimation of \({X}_i(t)\). Our overall objective is to iteratively update \(\{\hat{\pi}_{im}\}\) and \(\widehat{X}_i(t)\) until convergence, maximizing the likelihood for all subjects and their time segments.
We outline the iterative procedure for estimating both the latent functional covariates \(\{X_i(t)\}\) and the zero-inflation probabilities \(\{\pi_i(t)\}\) in Algorithm~\ref{alg:estimation}. 
We refer to this procedure as BE-ZIME, reflecting its use of basis expansions (BE) and its simultaneous correction for zero inflation (ZI) and measurement error (ME).

\begin{algorithm}
\caption{\enskip Iterative Estimation Procedure BE-ZIME}\label{alg:estimation}
\begin{algorithmic}
\State \textbf{Input:} Observations \(\{W_{ij}(t)\}\); basis functions \(\{\phi_k(t)\}_{k=1}^K\); initial estimates \(\{\hat{\pi}_i(t)\}\); and the assumed number of segments \(\widehat{M}\) and corresponding 
partition \(\{\widetilde{\mathcal{T}}_m\}_{m=1}^{\widehat{M}}\).

\State \textbf{Output:} Final estimates of the latent functional covariates \(\{\widehat{X}_i(t)\}\) and the zero-inflation probabilities \(\{\widehat{\pi}_i(t)\}\).

\vspace{0.3cm}
\noindent\textbf{Step 1. Adjust Zero Inflation and Perform Basis Expansion.} 
    For each subject \(i\), adjust \(\{W_{ij}(t)\}_{j=1}^J\) by the current working zero-inflation probability to get
    \begin{equation}\label{eq:wadj}
        \widehat{W}_{ij}^*(t)
        \;=\;
        \frac{W_{ij}(t)}{1-\hat{\pi}_i(t)}.
    \end{equation}
    Then project \(\widehat{W}_{ij}^*(t)\) onto a chosen set of basis functions \(\{\phi_k(t)\}_{k=1}^K\). For each subject \(i\), day \(j\), and basis index \(k\), let
    \begin{equation}\label{eq:winp}
        w_{ijk}^*
        \;=\;
        \int_0^1 
        \widehat{W}_{ij}^*(t)
        \,\phi_k(t)\,dt.
    \end{equation}

\noindent\textbf{Step 2. LMM for Measurement Error Correction.}
    {For each $k=1,\dots,K$, fit the LMM
    \begin{equation}\label{eq:lmm}
        w_{ijk}^*
        \;=\;
        \alpha_{0k}
        \;+\;
        \alpha_{ik}
        \;+\;
        \varepsilon_{ijk},
    \end{equation}
where $\alpha_{0k}$ is a fixed intercept across all subjects, $\alpha_{ik}$ is a subject-specific random intercept, and $\varepsilon_{ijk}$ is the residual error.
Let $\hat x_{ik} = \hat\alpha_{0k} + \hat\alpha_{ik}$ and $\hat{x}_i = (\hat x_{i1},\dots,\hat x_{iK})^\top$, and then reconstruct $\widehat X_i(t)$ as 
    \begin{equation}\label{eq:xhat}
        \widehat{X}_i(t)
        \;=\;
        \Phi(t)^\top
        \left\{
            \int_0^1 \Phi(s)\Phi(s)^\top ds
        \right\}^{-1}
        \hat{x}_i,
    \end{equation}
where $\Phi(t) = (\phi_1(t),\dots,\phi_K(t))^\top$ and $\int_0^1 \Phi(s)\Phi(s)^\top ds$ is a $K\times K$ matrix.
}

\vspace{0.2cm}
\noindent\textbf{Step 3. Update \(\pi_i(t)\) via Maximum Likelihood.} 
    With the updated \(\widehat{X}_i(t)\) from~\eqref{eq:xhat}, plug it into the likelihood and maximize 
    \[
        \ell_i^{(m)}\bigl\{\pi_{im}\mid \widehat{X}_i(t)\bigl\}
        \;=\;
        \prod_{t\in\widetilde{\mathcal{T}}_m}\;\prod_{j=1}^J\;
        \ell_{ij}\bigl\{t\mid \widehat{X}_i(t),\,\pi_i(t)=\pi_{im}\bigr\},
    \]
    to obtain
    \[
        \hat{\pi}_{im}
        \;=\;
        \arg\max_{\pi_{im}}
        \;\ell_i^{(m)}\bigl\{\pi_{im}\mid \widehat{X}_i(t)\bigl\}
    \]
    for $m=1,\dots,\widehat{M}$, and set \(\hat{\pi}_i(t) = \hat{\pi}_{im}\) for \(t\in\widetilde{\mathcal{T}}_m\). 

    \vspace{0.2cm}
\noindent\textbf{Step 4. Repeat Steps 1-3 Until Convergence.} 
    Repeat Steps 1--3 until numerical convergence, yielding final estimates \(\{\widehat{X}_i(t)\}\) and \(\{\widehat{\pi}_i(t)\}\).

\end{algorithmic}
\end{algorithm}

{
As shown in Algorithm~\ref{alg:estimation}, Step~1 adjusts each subject's measurements by the current working zero-inflation probability via \eqref{eq:wadj}, so that, under Assumptions~\ref{assump:decomp}--\ref{assump:indep}, $\mathbb{E}\{\widehat W_{ij}^*(t)\mid X_i(t)\}\approx X_i(t)$ when $\widehat\pi_i(t)\approx\pi_i(t)$.
We then project $\widehat W_{ij}^*(t)$ onto the chosen basis
$\{\phi_k(t)\}_{k=1}^K$ as in \eqref{eq:winp}, obtaining scalar scores $w_{ijk}^*$ for each $k$. 
For each fixed $k$, $\{w_{ijk}^*: j=1,\dots,J\}$ are repeated noisy projections for subject $i$ on the $k$th basis function, which we model using \eqref{eq:lmm}.
Step~2 thus leverages the within-subject replications to adjust for measurement error via linear mixed models \citep{laird1982random} (LMM).
Specifically, for each $k$ we fit \eqref{eq:lmm} and define
$\hat x_{ik} = \hat\alpha_{0k} + \hat\alpha_{ik}$ as an estimator of $x_{ik} := \int_0^1 X_i(t)\,\phi_k(t)\,dt$, since under our assumptions $E\{w_{ijk}^* \mid X_i\} = x_{ik}$.
Applying the least-squares reconstruction \eqref{eq:xhat} to
$\hat x_i = (\hat x_{i1},\dots,\hat x_{iK})^\top$ then yields the
corrected trajectory $\widehat X_i(t)$.
When the basis functions are orthonormal, \eqref{eq:xhat} simplifies to $\widehat X_i(t) = \Phi(t)^\top \hat x_i= \sum_{k=1}^K \hat x_{ik}\,\phi_k(t)$.
}
Step~3 updates \(\{\hat\pi_i(t)\}\) by maximizing the likelihood given the newly updated \(\widehat{X}_i(t)\). 
Finally, Step~4 repeats this process until convergence of \(\hat{\pi}_{i}(t)\); that is, when the average absolute change \(\sum_{m=1}^{\widehat{M}} |\hat{\pi}_{im}^{(r)} - \hat{\pi}_{im}^{(r-1)}| / \widehat{M}\) falls below a pre-specified threshold, where \(r\) indexes the iteration number. We used a threshold of \(10^{-3}\) in our experiments, and the algorithm typically converged within 10 iterations. 
This iterative procedure estimates subject-specific zero-inflation probabilities to account for structural zeros, while simultaneously correcting for measurement error.

An initial estimate of the zero-inflation probability is required to start the iterative routine. A natural yet naive choice is
\[
   \hat{\pi}_{im}
   \;=\;
   \frac{1}{a_{im} J}
   \sum_{t \in \widetilde{\mathcal{T}}_m}
   \sum_{j=1}^J
   \mathbb{I}\bigl\{\,W_{ij}(t)\,=\,0\bigr\},
\]
where \(a_{im}\) is the number of observed time points within the interval \(\widetilde{\mathcal{T}}_m\). Although this initial estimator is biased for \(\pi_i(t)\), our simulation studies confirm that the iterative scheme rapidly corrects this bias, resulting in an accurate estimate of \(\pi_i(t)\). Additionally, it is well known that model selection in measurement error settings is inherently challenging \citep{ma2010variable}. 
{For the choice of basis functions $\{\phi_k(t)\}_{k=1}^K$, common options include B-splines and Fourier bases.}
The number of basis functions $K$ can be selected using classical criteria that ignore measurement error and zero inflation, such as generalized cross-validation \citep{tekwe2019instrumental} or information criteria \citep{chen2024adjusting}. 
{Specifically, we choose the $K$ that minimizes the Bayesian information criterion (BIC) in the joint quantile regression fit, 
targeting goodness-of-fit for the second-stage model rather than exact reconstruction of $X_i(t)$.}
The number of segments \(M\) used to model the piecewise-constant zero-inflation probability is generally unknown and must be specified by the analyst. A practical strategy is to select \(M\) based on prior knowledge or exploratory data visualization. Additional discussion and empirical evaluations of how different choices of \(M\) affect estimation performance are provided in the simulation studies (Section~\ref{ssec:sim_pc}). The results suggest that when \(\widehat{M} \geq M\), the resulting performance is similar and generally better than those obtained when \(\widehat{M} < M\). This indicates that slightly overestimating \(M\) may be advantageous in practice, provided the density of observed time points is sufficient to support stable segment-specific estimation.

In the second stage, we implement functional quantile regression by substituting \(\widehat{X}_i(t)\) into~\eqref{eq:fqm}, thus replacing the integral \(\int_0^1 \beta(t,\tau)\,X_i(t)\,dt\) with \(\int_0^1 \beta(t,\tau)\,\widehat{X}_i(t)\,dt\). Standard approaches \citep{yao2017regularized,li2022inference} can then be applied to solve
\[
    \min_{\beta(\cdot,\tau),\,\btheta(\tau)}
    \sum_{i=1}^n 
    \rho_\tau\Bigl\{\,
      Y_i - \int_0^1 \beta(t,\tau)\,\widehat{X}_i(t)\,dt 
      - \bZ_i^T\,\btheta(\tau)
    \Bigr\},
\]
where \(\rho_\tau(u) = u \bigl[\tau - \mathbb{I}\{u<0\}\bigr]\) is the standard check loss. 
{In implementation, instead of working with $\widehat X_i(t)$ directly, we use the estimated vector $\hat x_i$ from Algorithm~\ref{alg:estimation} as scalar covariates.
We regress $Y_i$ on $(\hat x_i,\boldsymbol Z_i)$ and reconstruct $\hat\beta(t,\tau) = \sum_{k=1}^K \hat\gamma_k(\tau)\,\phi_k(t)$, where $\hat\gamma_k(\tau)$ is the estimated coefficient for $\hat x_{ik}$.
This reparameterization avoids repeated basis expansions of $\widehat X_i(t)$ and allows us to apply
standard joint quantile regression methods for scalar predictors\citep{muggeo2022additive}.
}
This procedure yields estimates \(\widehat{\beta}(t,\tau)\) and \(\widehat{\btheta}(\tau)\). Because \(\widehat{X}_i(t)\) accounts for both zero inflation and measurement error, the resulting quantile regression reliably captures how the underlying latent covariate \(X_i(t)\) influences the conditional quantiles of \(Y\). 
In addition, 
quantile estimates at multiple levels can be jointly estimated to improve numerical stability and enforce the monotonicity of quantile curves. 
In our implementation, 
we incorporate a joint quantile regression framework \citep{bondell2010noncrossing,hu2024simultaneous}, which is widely used in the literature to produce coherent and interpretable quantile estimates across levels.

In summary, the proposed approach first recovers \(\{\hat\pi_i(t)\}\) and \(\widehat{X}_i(t)\) via an iterative algorithm. This procedure accounts for zero inflation by estimating the probability of structural zeros and corrects measurement error. The recovered latent curves \(\widehat{X}_i(t)\) are then incorporated into model \eqref{eq:fqm} to capture how \(X_i(t)\) influences different parts of the distribution of \(Y\). This approach is computationally feasible and sufficiently flexible to handle intricate measurement error as well as subject-specific, time-varying zero-inflation probabilities.

\section{Simulation Studies}\label{sec:simu}
This section examines the finite sample performance of BE-ZIME, in comparison to several existing measurement error correction methods, all employing two-stage-based error correction approaches under various simulation settings. 
{For each competing method, we first obtain its estimator $\widehat X_i(t)$ as described below, compute $\hat x_{ik} = \int_0^1 \widehat X_i(t)\,\phi_k(t)\,dt$, and then apply the same stage-two joint quantile regression procedure outlined in Section~\ref{subsec:est-procedure} as for BE-ZIME.}
{All methods use the same cubic B-spline basis for functional expansion to ensure a fair comparison. Joint quantile regression is implemented via the function \texttt{gcrq} from the \texttt{quantregGrowth} \citep{muggeo2022additive} package, and separate quantile fits are obtained using \texttt{rq} from the \texttt{quantreg} \citep{koenker2018package} package.}
The competing approaches are described below:
\begin{itemize}
    \item \textbf{Naive}: Directly uses the first replicate \(W_{i1}(t)\) as a substitute for \(X_i(t)\).
    \item \textbf{Average}: Averages all replicates \(\{W_{ij}(t)\}_{j=1}^J\) to obtain \(\overline{W}_i(t)\), taken as \(\widehat{X}_i(t)\).
    \item \textbf{P-LMM} \citep{cui2022fast}: {Fits a linear mixed model at each time point $t$ to $W_{ij}(t)$ with a subject-specific random effect, and takes the fitted subject-specific mean trajectory as $\widehat X_i(t)$.}
    \item \textbf{P-PMM} and \textbf{P-ZIPMM}: {As in P-LMM, but using respectively a Poisson mixed model (PMM) and a zero-inflated Poisson mixed model (ZIPMM) at each time point $t$; the fitted subject-specific mean trajectories are used as $\widehat X_i(t)$.}
    \item \textbf{BE-ME}: {Applies the same basis-expansion and LMM procedure as BE-ZIME but without the zero-inflation adjustment, i.e., treating $\pi_i(t)\equiv 0$, to obtain $\widehat X_i(t)$.}
    \item \textbf{Oracle}: {Uses the true functional covariate $X_i(t)$ as $\widehat X_i(t)$, serving as a benchmark.}   
\end{itemize}
Prior to presenting empirical performance through simulation, we briefly discuss the methodological assumptions and differences among the competing approaches to clarify their expected performance and potential limitations.
The Naive approach overlooks both measurement error and excess zeros, resulting in biased recovery of the latent process and failing to leverage repeated measurements. The Average estimator reduces random variability caused by measurement error by aggregating replicates, but it does not account for structured excess zeros. The pointwise mixed-model methods, P-LMM and P-PMM, and the basis-expansion measurement-error correction method (BE-ME) apply mixed models to address measurement error, yet fail to account for zero inflation. Specifically, the normality assumption of P-LMM is inappropriate for modeling nonnegative count data, and both P-PMM and BE-ME assume that the zero-inflation probability is zero. In contrast, P-ZIPMM and our proposed method BE-ZIME correct both measurement error and zero inflation.

{Throughout our simulations, data are generated independently from a functional quantile regression model in which, 
\begin{align*}
    Q_{Y_i}(\tau \mid X_i, \bZ_i) &= \int_0^1 \beta(t,\tau)\,X_i(t)\,dt \;+\;Z_{1i}\theta_1(\tau) \;+\; Z_{2i}\theta_2(\tau) \;+\; F_\varepsilon^{-1}(\tau),\\
    W_{ij}(t) \; &\sim\;\mathrm{ZIP}\bigl\{\,X_i(t)\,\bigm|\;\pi_i(t)\bigr\},\quad j=1,\dots,J,\quad i=1,\dots,n,\label{eq:W-def},
\end{align*}
where $\beta(t,\tau)=\{1+\eta_\beta h(\tau)\}\,\beta_0(t)$, $F_{\varepsilon}^{-1}(\tau)=\sigma\{1+\eta_\varepsilon h(\tau)\}\,\Phi^{-1}(\tau)$ denotes the $\tau$th quantile of the regression error $\varepsilon_i$, 
$h(\tau)=0.2(\tau-0.5)$, $\beta_0(t)$ is the baseline shape, $(\eta_\beta,\eta_\varepsilon)$ control the strengths of coefficient and error heterogeneity, respectively, and $\Phi^{-1}$ is the standard normal quantile function.
}
The functional covariate is generated as $X_i(t) = \sum_{k=1}^{50}\xi_{ik}\zeta_k\phi_k(t) + 5$, with \(\xi_{ik}\) i.i.d.\ from \(\mathrm{Uniform}(-\sqrt{3}, \sqrt{3})\) and \(\zeta_k=(-1)^{k+1}k^{-1}\). The basis functions are defined by $\phi_1(t)=1$ and $\phi_k(t) = \sqrt2\cos((k-1)\pi t)$ for $k \geq 2$. The surrogate \(W_{ij}(t)\) arises from a zero-inflated Poisson (ZIP) distribution with mean \(X_i(t)\) and zero-inflation probability \(\pi_i(t)\). For each subject \(i\), we record \(L\) equally spaced observations of \(W_{ij}(t)\) on \([0,1]\). 
We also include two error-free scalar covariates, \(Z_{1i}\) and \(Z_{2i}\), drawn independently as \(Z_{1i} \sim \mathcal{N}(1,\sigma_z^2)\) and \(Z_{2i}\sim\mathrm{Bernoulli}(p_z)\).
In all scenarios, we set the scalar coefficients at $\theta_1(\tau) \equiv \theta_1 = 0.5$ and $\theta_2(\tau) \equiv \theta_2 = 0.6$ and the parameters $\sigma_z = 0.5$ and $p_z = 0.6$.

{
We consider two simulation regimes. In Section~\ref{ssec:sim_cons}, we focus on a baseline setting with subject-specific but time-constant zero-inflation probabilities $\pi_i(t)\equiv\pi_i$, homoscedastic errors ($\eta_\varepsilon=0$), and a quantile-invariant coefficient $\beta(t,\tau)\equiv\beta(t)$. 
This controlled design helps us to clearly assess how zero inflation and basic design factors affect the performance of the competing methods.
In Section~\ref{ssec:sim_pc}, we move to a more realistic scenario with piecewise-constant $\pi_i(t)$, quantile-varying functional effects $\beta(t,\tau)$, and heteroscedastic errors across quantile levels, to evaluate the robustness of BE-ZIME and its competitors under complex zero-inflation and heterogeneity patterns.
}

To assess model performance across the outcome distribution, we consider quantile levels \(\tau\in\{0.25,0.50,0.75\}\). We conduct simulations for sample sizes \(n\in\{100,500\}\).
We evaluate the estimation accuracy of the functional coefficient \(\beta(t,\tau)\) in our joint quantile regression framework by computing the average squared bias (\(\mathrm{ABias}^2\)), average sample variance (\(\mathrm{AVar}\)), and mean integrated squared error (\(\mathrm{MISE}\)):
\begin{align*}
    \mathrm{ABias}^2\left(\widehat{\beta}_\tau\right) &= \frac{1}{L} \sum_{\ell = 1}^{L} \left\{ \overline{\beta}(t_\ell,\tau) - \beta(t_\ell,\tau) \right\}^2,\\
    \mathrm{AVar}\left(\widehat{\beta}_\tau\right) &= \frac{1}{RL} \sum_{r=1}^R \sum_{\ell = 1}^{L} \left\{ \widehat{\beta}^{(r)}(t_\ell,\tau) - \overline{\beta}(t_\ell,\tau) \right\}^2,\\
    \mathrm{MISE}\left(\widehat{\beta}_\tau\right) &= \text{ABias}^2 + \text{Avar},
\end{align*}
where \(R=500\) is the number of simulation replicates, \(L\) is the number of discrete time points in \([0,1]\), \(\widehat{\beta}^{(r)}(t_\ell,\tau)\) denotes the estimated coefficient function at replication \(r\), and \(\overline{\beta}(t_\ell,\tau)\) is the corresponding pointwise mean across all replicates.

\subsection{Constant Zero-Inflation Probability}\label{ssec:sim_cons}

{
In this subsection, we adopt a baseline setting with $\eta_\beta=\eta_\varepsilon=0$ and $\sigma=0.1$, so $\beta(t,\tau)\equiv\beta(t)$ and $F^{-1}_\varepsilon(\tau)=0.1\,\Phi^{-1}(\tau)$ (equivalently, $\varepsilon_i \stackrel{\text{iid}}{\sim} N(0,0.1^2)$). The outcome is generated as $Y_i=\int_0^1 \beta(t)X_i(t)\,dt+\theta_1 Z_{1i}+\theta_2 Z_{2i}+\varepsilon_i$. 
We consider the functional coefficient to be either a smoother signal $\beta(t)=0.5\sin(\pi t)$ or a sharper signal $\beta(t)=\sin(2\pi t)$, and assume a constant zero-inflation regime with $\pi_i(t)\equiv\pi_i$, i.e., the zero-inflation probability is constant over time.}
To introduce subject-level heterogeneity in zero-inflation probabilities, we let
\[
    \pi_i \;\sim\;\mathrm{Uniform}\bigl(\pi_0-\pi_\delta,\;\pi_0+\pi_\delta\bigr),
\]
with \(\pi_0\) denoting the average zero-inflation probability and \(\pi_\delta\) controlling its variability among subjects. We consider:
\begin{description}
    \item[Case 1:] \(\pi_0=0, \pi_\delta=0\), i.e.\ $\pi_i=0$ for all $i$, indicating no zero inflation in the data.
    \item[Case 2:] \(\pi_0=0.3, \pi_\delta\in\{0,0.1,0.2\}\), representing a moderate level of zero inflation.
    \item[Case 3:] \(\pi_0=0.6, \pi_\delta\in\{0,0.1,0.2\}\), representing a high level of zero inflation.
\end{description}

{
We begin by evaluating recovery of the latent trajectories \(X_i(t)\) across all methods.
For replication \(b\), we compute the MISE for $\widehat X$ as \(\mathrm{MISE}_b(\widehat X)=(nL)^{-1}\sum_{i=1}^n\sum_{\ell = 1}^L\{\widehat X_i^{(b)}(t_\ell)-X_i^{(b)}(t_\ell)\}^2\),
and report the mean (and standard deviation) over \(R=500\) replications with \(n=L=100\) in Table~\ref{tab:s1xi}.
When there is no zero inflation \((\pi_0=\pi_\delta=0)\), BE-ZIME and BE-ME perform similarly, as expected.
As the overall zero-inflation level \((\pi_0)\) increases, BE-ZIME yields the lowest errors, whereas the remaining approaches show substantially higher errors.
}
We then evaluate the performance of our method in estimating the subject-specific zero-inflation probabilities \(\pi_i\) under the simulation settings considered in this subsection. We computed the mean squared error (MSE) as \(\text{MSE}_b(\hat{\pi}) = \frac{1}{n} \sum_{i=1}^n (\hat{\pi}_i^{(b)} - \pi_i)^2\) for each replicate \(b = 1, \dots, 500\). Table~\ref{tab:s1pi} reports the average and standard deviation (in parentheses) of these \(\text{MSE}_b(\hat{\pi})\) values across replications, under various combinations of zero-inflation levels, sample sizes \(n\), and numbers of observed time points \(L\). 
Across all settings, the estimation errors remain consistently small, indicating high accuracy of our method. The results suggest that increasing the zero-inflation level has only a mild effect on the estimation error. In contrast, increasing the number of observed time points from \(L = 100\) to \(L = 200\) leads to a noticeable reduction in error. Since we estimate subject-specific \(\pi_i\), increasing the sample size \(n\) does not significantly impact the estimation accuracy, which aligns with our model's structure.

\begin{figure}
\centering
\includegraphics[width=1\textwidth]{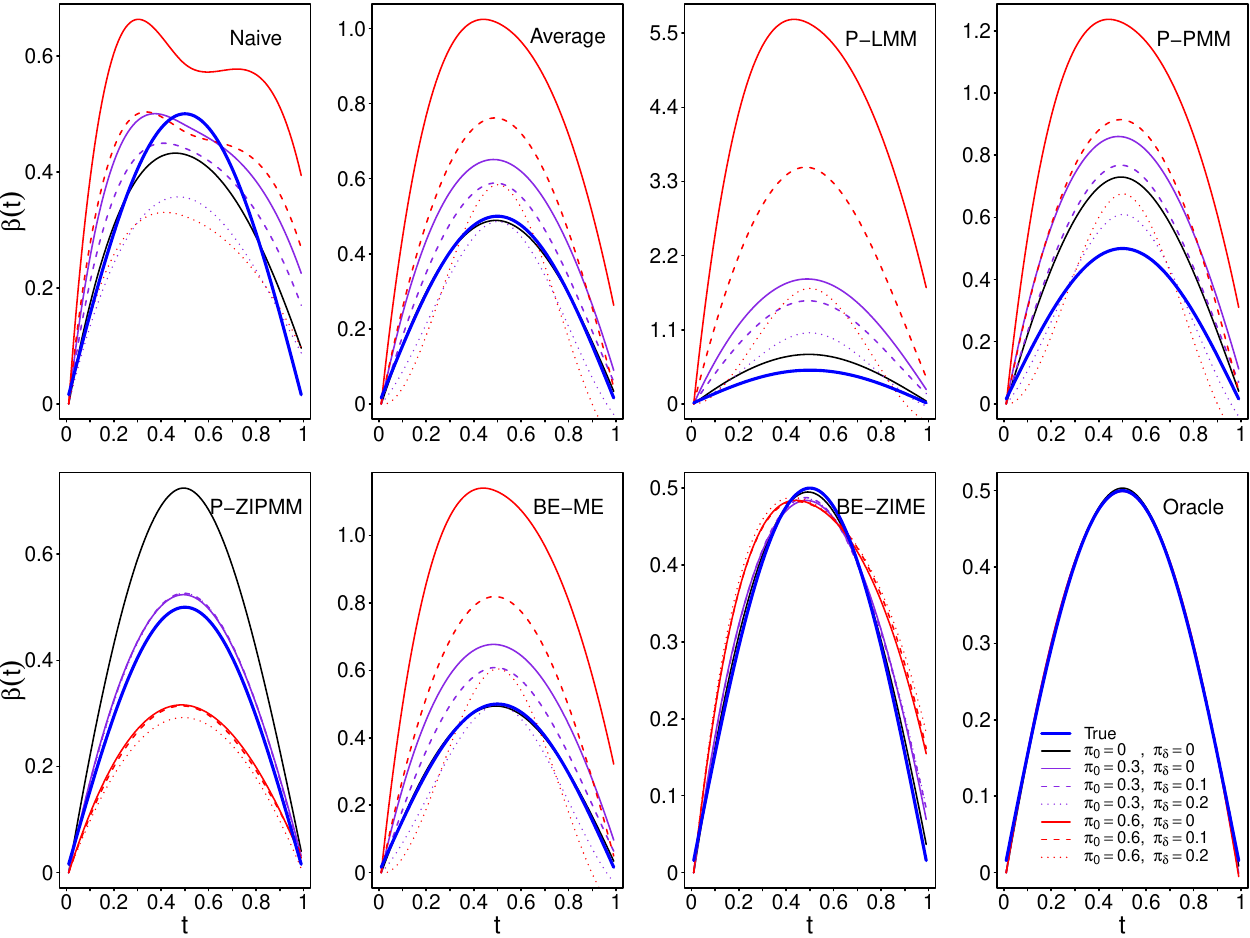}
\caption{Mean estimates of \(\beta(t)=0.5\sin(\pi t)\) at \(\tau=0.5\) under different zero-inflation levels, with sample size \(n = 100\) and observed time points \(L=100\), over 500 replications. Here, \(\pi_i\) represents the zero-inflation probability of the \(i\)-th subject, with \(\pi_i \sim \mathrm{Uniform}(\pi_0 - \pi_\delta,\, \pi_0 + \pi_\delta)\). The bold blue curve denotes the true coefficient.}
\label{fig:s1beta1}
\end{figure}

\begin{figure}
\centering
\includegraphics[width=1\textwidth]{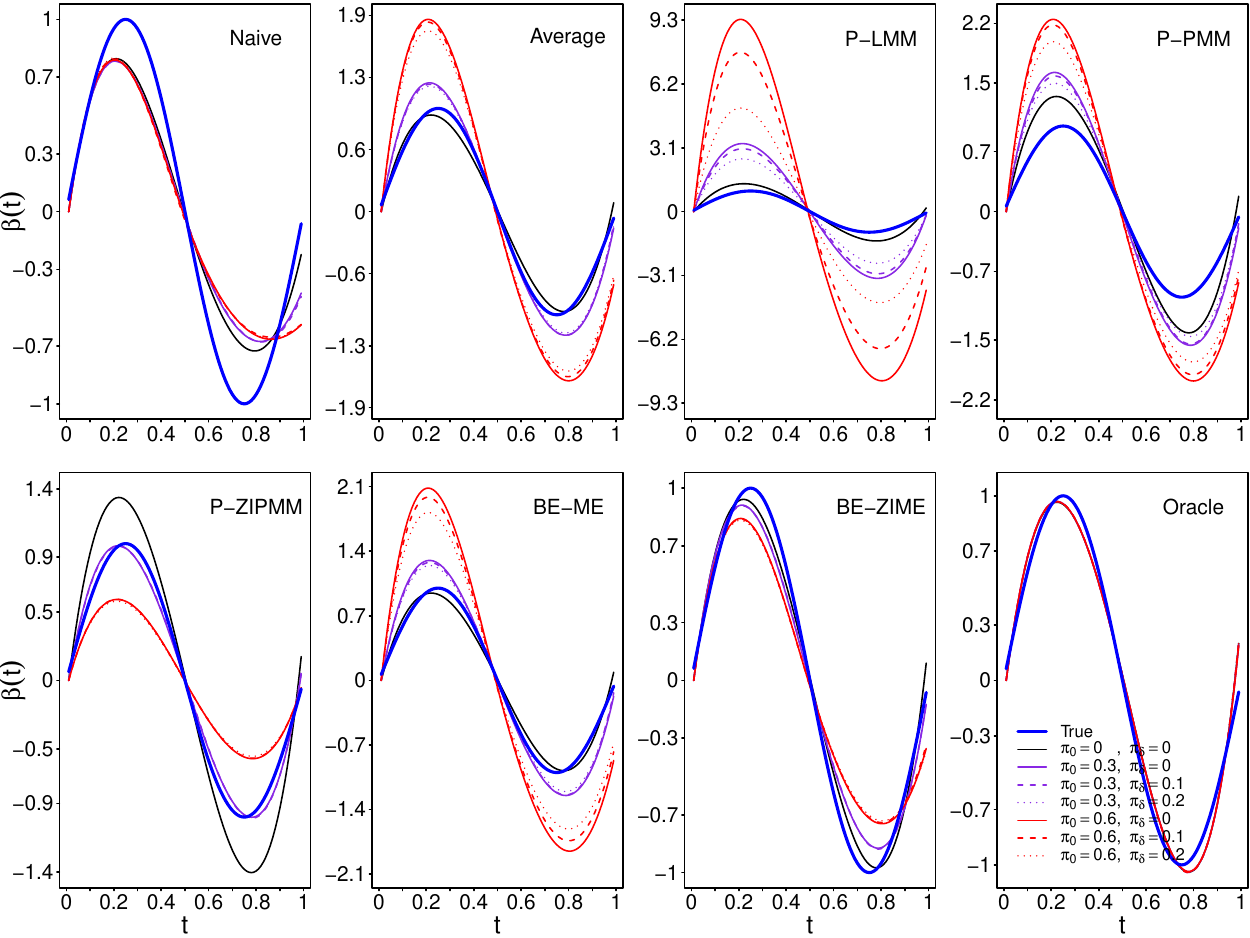}
\caption{Mean estimates of \(\beta(t)=\sin(2\pi t)\) at \(\tau=0.5\) under different zero-inflation levels, with sample size \(n = 100\) and observed time points \(L=100\), over 500 replications. Here, \(\pi_i\) represents the zero-inflation probability of the \(i\)-th subject, with \(\pi_i \sim \mathrm{Uniform}(\pi_0 - \pi_\delta,\, \pi_0 + \pi_\delta)\). The bold blue curve denotes the true coefficient.}
\label{fig:s1beta2}
\end{figure}

Figures~\ref{fig:s1beta1} and~\ref{fig:s1beta2} illustrate the mean estimated coefficient functions at the \(50\%\) quantile level averaged over \(R=500\) replications. In each plot, the true coefficient is represented by the bold blue line, and the vertical gap between each estimated line and the true line indicates the estimation bias. Across different zero-inflation levels, the proposed BE-ZIME consistently achieves the smallest deviations from the true curve and demonstrates mild sensitivity to increasing values of \(\pi_0\) or shifts in the variation parameter \(\pi_\delta\). Conversely, the BE-ME and Average methods are accurate when there is no zero inflation (\(\pi_0=0,\pi_\delta=0\)), however, their accuracy diminishes as \(\pi_0\) decreases. Among the pointwise-based methods, P-ZIPMM performs well under moderate zero inflation but falters under situations of no inflation or high inflation, possibly due to challenges in accurately estimating the zero-inflation probability. P-LMM, P-PMM, and Naive all demonstrated poor performance across all levels of zero inflation considered. In particular, the relatively poor performance of P-PMM at $\pi_0 = 0$ can be attributed to its inability to borrow information across time.

In Tables~\ref{tab:s1beta1n100} and~\ref{tab:s1beta1n500}, we present the $\mathrm{ABias}^2$, $\mathrm{AVar}$, and $\mathrm{MISE}$ for $\beta(t) = 0.5\sin(\pi t)$ at the $50\%$ quantile, using $L=100$ observations and sample sizes $n=100$ and $n=500$. When there's no zero inflation (i.e., $\pi_0=0, \pi_\delta=0$), the estimation bias and variance results are comparable across the Average, BE-ME, and the new BE-ZIME method. However, as zero inflation increases ($\pi_0 >0$), BE-ZIME attains the smallest MISE, approaching the Oracle's performance level, and its variance decreases markedly when $n$ increases from 100 to 500. Although P-ZIPMM achieves the smallest estimation variance under severe zero inflation ($\pi_0=0.6$), it exhibits a much larger bias than our proposed method, resulting in a higher MISE. Additionally, while Average, P-LMM, P-PMM, and BE-ME sometimes show lower $\mathrm{ABias}^2$ as $\pi_\delta$ increases (possibly due to a greater proportion of subjects exhibiting less severe zero-inflation, thereby reducing overall bias), their variances tend to spike in these scenarios, indicating instability and lack of robustness. Moreover, Naive occasionally outperforms Average under zero inflation, which differs from scenarios with only measurement error\citep{chen2024adjusting}. This may be because the Naive method relies on a single observation per subject, $W_{i1}(t)$, which has a higher chance of being unaffected by zero inflation compared to the average of multiple replicates $\overline{W}_i(t)$. These findings are consistent across other quantile levels (see Supplementary Section S1.1).

Tables~\ref{tab:s1beta2gp100} and~\ref{tab:s1beta2gp200} summarize analogous results for the sharper coefficient function \(\beta(t)=\sin(2\pi t)\), again at the \(50\%\) quantile, with \(n=100\), \(L=100\) and $200$, respectively. BE-ZIME continues to perform well overall, with \(\mathrm{ABias}^2\) and \(\mathrm{AVar}\) both improving considerably when \(L\) increases from 100 to 200. Under moderate zero inflation (\(\pi_0=0.3\)), however, P-ZIPMM can slightly outperform BE-ZIME for this sharper function. Nevertheless, BE-ZIME narrows the gap and nearly matches P-ZIPMM once \(L=200\) provides more density observations. One possible explanation is that denser observations of \(\{W_{ij}(t)\}\) yield a greater number of informative nonzero data points, facilitating more accurate curve recovery. Results for other quantile levels again follow the same patterns (see Supplementary Section S1.1).

To quantify the advantages of fitting multiple quantiles jointly rather than separately, we compute the ratio
\[
  \frac{\mathrm{MISE}_{\text{separate}} - \mathrm{MISE}_{\text{joint}}}
       {\mathrm{MISE}_{\text{separate}}},
\]
where \(\mathrm{MISE}_{\text{joint}}\) is the mean integrated squared error of the estimated functional coefficient under joint quantile regression, and \(\mathrm{MISE}_{\text{separate}}\) is the corresponding error when fitting each quantile level independently. A larger ratio indicates greater improvement achieved by joint estimation.
Tables~\ref{tab:s1jsbeta1} and~\ref{tab:s1jsbeta2} report these ratios at the $25\%$ and $75\%$ quantiles for \(\beta(t)=0.5\sin(\pi t)\) and \(\beta(t)=\sin(2\pi t)\), respectively, with \(L=100\) and \(n=100\). The ratio at the $50\%$ quantile is zero and therefore not included in the tables. In almost all cases, joint estimation yields substantial improvements compared to fitting separate quantile regressions.

\subsection{Piecewise Constant Zero-Inflation Probability}\label{ssec:sim_pc}

In this subsection, we consider a regime that allows quantile-varying functional effects and heteroscedastic errors.
For each subject $i$, we draw a latent quantile $U_i\sim\mathrm{Unif}(0,1)$ and set $\beta_i(t)=\beta(t,U_i)$ and $\varepsilon_i=\varepsilon(U_i)$, so that the outcome is $Y_i=\int_0^1 \beta_i(t)X_i(t)\,dt+\theta_1Z_{1i}+\theta_2Z_{2i}+\varepsilon_i$. We use the baseline shape $\beta_0(t)=0.5\sin(\pi t)$ and vary heterogeneity by setting $\eta_\beta,\eta_\varepsilon\in\{0,1\}$.

We now specify the piecewise-constant zero-inflation probability $\pi_i(t)$ to study the impact of the segment pattern and possible misspecification of the working partition.
We consider three segmentation designs:
\begin{description}
  \item[Case 1 (2 equal segments):]
  set $\pi_i(t) = \pi_0$ for $t \in [0,0.5)$ and $\pi_i(t) = 1-\pi_0$ for $t \in [0.5,1]$.
  
  \item[Case 2 (5 equal segments):] set
  $\pi_i(t)=\pi_0$ on $[0,0.2)\cup[0.4,0.6)\cup[0.8,1]$ and
  $\pi_i(t)=1-\pi_0$ on $[0.2,0.4)\cup[0.6,0.8)$.

  \item[Case 3 (3 unequal segments):] set
  $\pi_i(t)=\pi_0$ on $[0,0.1)\cup[0.6,1]$ and
  $\pi_i(t)=1-\pi_0$ on $[0.1,0.6)$.
\end{description}
These designs vary the alternation frequency of zero inflation (low in Case 1, high in Case 2) and the regularity of segment lengths (equal in Cases 1 and 2, unequal in Case 3). 
We take $\pi_0\in\{0.8,0.6\}$ control the inter-segment contrast in zero-inflation probabilities. 
For each setting, we fit BE-ZIME with working segment counts $\widehat M\in\{1,2,5,10\}$, partitioning $[0,1]$ into $\widehat M$ equal-length subintervals.

We first evaluate recovery of the latent trajectories \(X_i(t)\) across all methods. 
In each replication \(b\), we compute $\mathrm{MISE}_b(\widehat X)$ and report the mean (and standard deviation) over \(R=500\) replications for \(n\in\{100,500\}\) and \(L=100\) in Table~\ref{tab:s2xi}.
BE-ZIME has the smallest errors whenever the working segment count allows time variation in zero inflation (\(\widehat M\ge2\)). 
The misspecified \(\widehat M=1\) performs worst among the BE-ZIME variants. 
Increasing \(\widehat M\) from \(1\) to \(2,5,10\) substantially reduces MISE, with diminishing gains once \(\widehat M\) is at or above the true segmentation. 
Greater inter-segment contrast (\(\pi_0=0.8\) vs.\ \(0.6\))  increases errors for all methods.
Increasing the sample size from \(n=100\) to \(n=500\) leaves mean MISE largely unchanged but reduces Monte Carlo variability.

We then evaluate the performance of our method in estimating the subject-specific, piecewise-constant zero-inflation probabilities \(\pi_i(t)\) under the simulation settings considered in this subsection. For each \(\widehat{M}\), we define the corresponding equal-length partition \(\{\widetilde{\mathcal{T}}_m\}_{m=1}^{\widehat{M}}\) and compute \(\tilde{\pi}_{im} = \int_{\widetilde{\mathcal{T}}_m} \pi_i(t) dt\), setting \(\tilde{\pi}_i(t) = \tilde{\pi}_{im}\) for \(t \in \widetilde{\mathcal{T}}_m\). 
To assess performance, we compute two versions of MSE for each replicate \(b = 1, \dots, 500\): $\text{MSE}_b(\hat{\pi}) = \int (\hat{\pi}_i(t) - \pi_i(t))^2 dt$ and $\text{MSE}_b^\dagger(\hat{\pi}) = \int (\hat{\pi}_i(t) - \tilde{\pi}_i(t))^2 dt$.
The first error, \(\text{MSE}_b(\hat{\pi})\), measures the deviation from the true zero-inflation structure, while the second, \(\text{MSE}_b^\dagger(\hat{\pi})\), reflects how well the method recovers the average zero-inflation probability over the assumed segments \(\{\widetilde{\mathcal{T}}_m\}\). 

Table~\ref{tab:s2pi} presents the averages and standard deviation (in parentheses) of both MSEs across replications for various cases and the assumed number \(\widehat{M}\).
{The $\text{MSE}_b(\hat{\pi})$ results show that estimation error is smallest when $\widehat M$ is the smallest segment count for which the working zero-inflation pattern matches the truth (i.e., $\tilde{\pi}_i(t)=\pi_i(t)$); in the equal-segment designs, this occurs when $\widehat M=M$.}
Notably, cases where \(\widehat{M}\) is an integer multiple of \(M\) also yield relatively low errors, although they exhibit slightly larger variability. This is expected, as a larger \(\widehat{M}\) reduces the number of time points per segment, increasing the variability in estimating \(\pi_{im}\). 
Additionally, settings with \(\pi_0 = 0.8\), indicating larger differences in zero-inflation probabilities across adjacent segments, make the model more sensitive to mis-specification of \(\widehat{M}\).
The results based on \(\text{MSE}_b^\dagger(\hat{\pi})\) confirm that our method accurately estimates the average zero-inflation probability over the assumed segments \(\{\widetilde{\mathcal{T}}_m\}\). 
Note that while smaller values of $\widehat{M}$ result in lower \(\text{MSE}_b^\dagger(\hat{\pi})\), this does not necessarily imply that a smaller $\widehat{M}$ is better. This is because a smaller $\widehat{M}$ means each segment contains more observations, which contributes to more stable estimation.

{
Tables~\ref{tab:s2beta1eta00} and \ref{tab:s2beta1eta11} summarize $\mathrm{ABias}^2$, $\mathrm{AVar}$, and $\mathrm{MISE}$ at the $50\%$ quantile for $\beta(t,\tau)$ with $n\in\{100,500\}$, under a quantile-invariant, homoscedastic specification $(\eta_\beta,\eta_\varepsilon)=(0,0)$, and a quantile-varying, heteroscedastic specification $(\eta_\beta,\eta_\varepsilon)=(1,1)$, respectively.
In all settings, the estimation variance decreases substantially as $n$ increases from 100 to 500, and the smaller-jump scenario $\pi_0=0.6$ yields better performance over the larger-jump case $\pi_0=0.8$. Moreover, both the Average and BE-ME approaches, which ignore zero inflation, perform poorly across all scenarios. Although P-ZIPMM achieves the smallest estimation variance, it suffers from substantially higher bias, leading BE-ZIME-$\widehat{M}$ to excel in terms of MISE, especially in the $\pi_0=0.8$ setting. 
For our proposed method, increasing $\widehat{M}$ typically reduces bias but increases variance. 
In most cases, performance stabilizes once the working segmentation is flexible enough to capture the main changes in the zero-inflation pattern. Under-segmentation yields noticeably larger MISE. This pattern aligns with the behavior observed for $\widehat{X}_i$ and $\widehat{\pi}_i$, and suggests that choosing a slightly larger $\widehat{M}$ can be advantageous in practice, provided the sample size is sufficient to support stable segment-specific estimation.
Finally, when we compare the homogeneous case $(\eta_\beta,\eta_\varepsilon)=(0,0)$ with the heterogeneous case $(\eta_\beta,\eta_\varepsilon)=(1,1)$, $\mathrm{ABias}^2$ remains similar while $\mathrm{AVar}$ increases under $(1,1)$, as expected.
Results at other quantiles and for $(\eta_\beta,\eta_\varepsilon)=(0,1)$ or $(1,0)$ show the same patterns (Supplementary Section~S1.2).
}

In summary, both simulation studies (constant and piecewise-constant ZI) confirm that BE-ZIME effectively accounts for measurement error and zero inflation while simultaneously modeling multiple quantiles. 
{The method maintains low estimation error and is stable across homogeneous and heterogeneous settings and a range of zero-inflation patterns, consistently outperforming existing methods and closely tracking the Oracle benchmark.}

\section{Real Data Application}\label{sec:real}

We illustrate our proposed methodology using data from a childhood obesity study conducted over an 18-month period by Dr. Mark Benden and colleagues, approved by the Institutional Review Board (IRB) Human Subjects Protection Program at Texas A\&M University \citep{benden2014evaluation}. In this study, energy expenditure (EE) and step counts (SC) were collected per minute via the SenseWear Armband\textsuperscript{\textregistered} (BodyMedia, Pittsburgh, PA) from 374 children who wore accelerometers during school hours for one week (five days). Both EE and SC observations are subject to measurement error, with their true values representing latent EE and SC, respectively \citep{tekwe2019instrumental,zhang2023partially}. Children's body weight, height, age, and sex were recorded, and their BMIs were calculated at the beginning of each semester throughout the study period. These variables serve as error-free scalar covariates in the analysis.

In this section, we aim to apply our proposed method to investigate the relationship between physical activity levels and the change in BMI over the 18 months. EE measures the calories or energy expended to support bodily functions, while SC reflects the number of steps taken over a specific period of time. Both are commonly used to assess physical activity levels. Although EE is often considered a more accurate indicator, it is generally more challenging to collect in practice. As noted by Benden et al \cite{benden2014evaluation}, EE measurements are highly sensitive to environmental conditions and device placement, require consistent skin contact, and need individualized calibration based on demographic and physiological variables. These factors introduce logistical complexity and elevate risks related to data quality, making routine EE collection challenging in large-scale or longitudinal studies. Furthermore, accurate EE-measuring devices like multi-sensor armbands \citep{koehler2017monitoring} are expensive and less accessible outside research settings. In contrast, SC is easily collected through standard accelerometry, requires no individualized calibration, and can be captured reliably using inexpensive devices such as wrist-worn fitness trackers or smartphones \citep{henriksen2018using}. However, SC data typically include many zeros, leading to a zero-inflation issue.

If SC-based regression coefficients closely match those derived from EE after addressing zero inflation and measurement error, SC could serve as a practical and scalable alternative measure of physical activity in resource-limited scenarios. Both SC and EE data availability in the childhood obesity study provides an ideal setting to test this hypothesis. Specifically, this study aims to
\begin{enumerate}
    \item Correct the measurement error and zero inflation in SC and the measurement error in EE (Section~\ref{ssec:real_proc}).
    \item Scale the corrected SC and EE measures for direct regression coefficient comparison (Section~\ref{ssec:real_proc}).
    \item Fit scalar-on-function quantile regression models separately using these scaled and corrected measures (SC vs.\ EE) to compare the resulting estimated functional coefficients (Section~\ref{ssec:real_est}).
\end{enumerate}

\subsection{Data Processing and Visualization}\label{ssec:real_proc}

\begin{figure}
\centering
\includegraphics[width=1\textwidth]{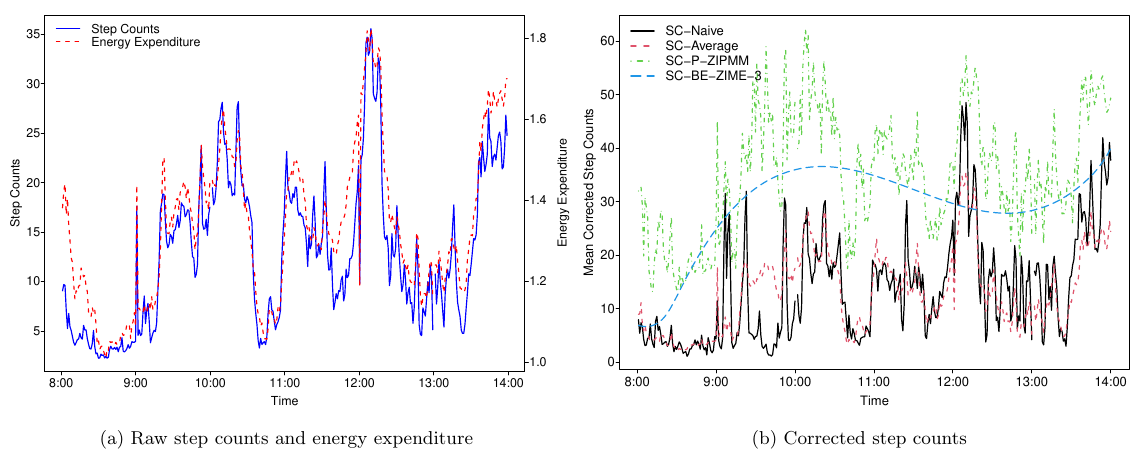}
\caption{Visualization of functional covariates. (a) Mean minute-level step counts and energy expenditure between 8:00 AM and 2:00 PM on separate axes, aggregated across students; (b)Mean corrected step counts estimated using four different methods over the same time period.}
\label{fig:SCcurves}
\end{figure}

Students in this study attended three elementary schools within the College Station Independent School District. They were randomly assigned either to stand-biased desks (\(\text{Treatment}=1\)) or traditional desks (\(\text{Treatment}=0\)) \citep{benden2011impact}. Scalar variables recorded included age, gender (girl=1, boy=0), and ethnicity (white=1, other=0). Students' schools were encoded via two dummy variables (\(\text{School}_k = 1\) if the student is from the \(k\)th school, \(k = 2,3\)). In addition to these variables, we also have data on teacher and grade within each school. However, previous studies have shown that these variables do not significantly influence BMI \citep{tekwe2019instrumental,honvoh2024modeling}. Therefore, they are not included in our analysis here. Students with missing BMI or covariates were excluded, yielding a final sample of \(n=173\). Table~\ref{tab:real_cov} summarizes the key error-free scalar variables. For the physical activity data, EE and SC were measured per minute from 8:00 AM to 2:00 PM over five days. Figure~\ref{fig:SCcurves}(a) depicts mean minute-level SC and EE aggregated across subjects on separate axes, revealing a synchronized pattern indicative of a near-linear relationship between SC and EE.

\begin{figure}
\centering
\includegraphics[width=\textwidth]{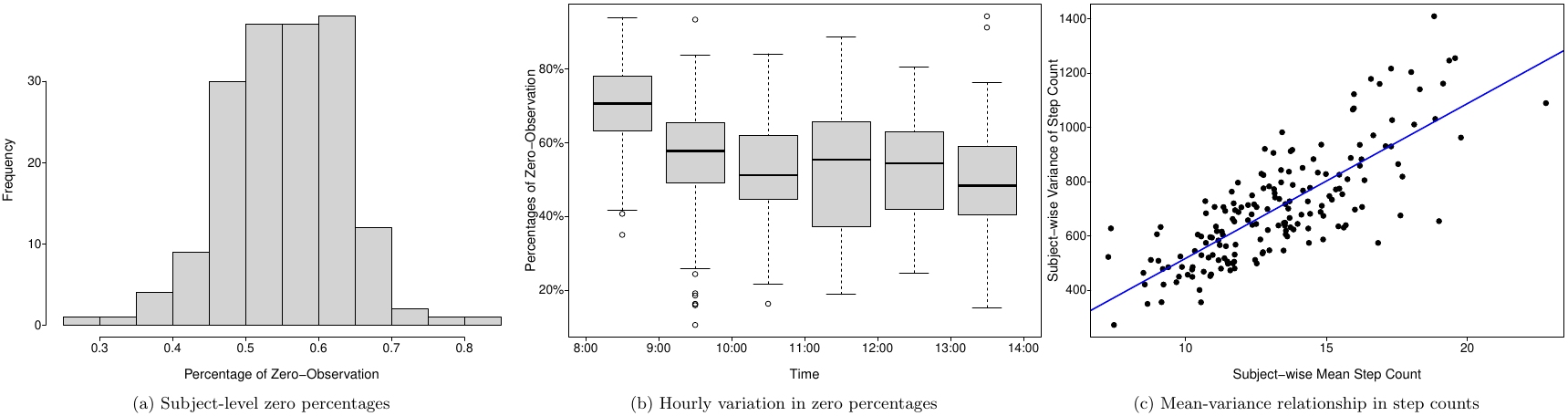}
\caption{Analysis of zero-inflated patterns in children's step counts. (a) Histogram showing the distribution of zero-observation percentages among students, reflecting strong subject-specific heterogeneity; (b) Box plots of zero-observation percentages across hour intervals, illustrating substantial time variation in zero frequencies; (c) Mean–variance scatter plot across all time points, with each point representing one student, reveals a strong linear relationship and substantial overdispersion, supporting the presence of zero inflation. The fitted linear model is: $\text{Variance} = -53.25 + 57.02 \times \text{Mean}$.}
\label{fig:zeroobs}
\end{figure}

The SC observations include a substantial proportion of zeros potentially due to inactivity, such as extended periods of siting, or device-related issues such as non-wear or incomplete recordings. 
Figure~\ref{fig:zeroobs}(a) presents a histogram of the percentage of zero observations per student, ranging from 25.6\% to 83.0\% with a mean around 55\%, indicating considerable subject-specific heterogeneity. A standard Poisson model would inadequately capture this distribution due to the prevalence of zeros, motivating the use of a ZIP model. 
Figure~\ref{fig:zeroobs}(b) displays box plots of zero percentages across hourly intervals from 8:00 AM to 2:00 PM, revealing systematic temporal variation in zero frequencies. Figure~\ref{fig:zeroobs}(c) depicts the mean–variance relationship of step counts across school hours. A clear positive linear trend is observed, with variance substantially exceeding the mean, indicating strong overdispersion. These results collectively motivate the need to account for subject-specific and time-varying zero inflation. Additional hour-level analyses are provided in Supplementary Section S2.

In this application, we denote by $X_{\text{EE},i}(t)$ and $X_{\text{SC},i}(t)$ the true latent EE and SC, respectively, for student $i$. However, in practice, we observe only replicated measurements based on devices, indicated by $W_{\text{EE},ij}(t)$ and $W_{\text{SC},ij}(t)$, where $j=1,\dots,5$. Specifically, $W_{\text{EE},ij}(t)$ is subject to measurement error, while $W_{\text{SC},ij}(t)$ is affected by both measurement error and zero inflation.
Before fitting the regression model, we apply several correction methods to account for these measurement error and zero-inflation issues.  Specifically, the methods considered include: 

\begin{itemize}
    \item \textbf{SC-Naive:} Directly uses the first replicate \(W_{\text{SC},i1}(t)\) as a substitute for \(X_{\text{SC},i}(t)\).
    \item \textbf{SC-Average:} Averages all replicates \(\{W_{\text{SC},ij}(t)\}_{j=1}^5\) to obtain \(\overline{W}_{\text{SC},i}(t)\), taken as \(\widehat{X}_{\text{SC},i}(t)\).
    \item \textbf{SC-P-ZIPMM:} Fits a ZIPMM at each time point \(t\), to estimate $X_{\text{SC},i}(t)$.
    \item \textbf{SC-BE-ZIME-\(\widehat{M}\):} Our proposed correction method, employing a basis expansion combined with a piecewise-constant zero-inflation model segmented into \(\widehat{M}\) intervals.
    \item \textbf{EE-P-LMM:} Fits a LMM at each time point \(t\) to estimate \(X_{\text{EE},i}(t)\).
\end{itemize}
For our proposed method, we select \(\widehat{M} = 3\) to achieve a good balance between flexibility and estimation stability. As a benchmark for comparison, we use the EE-P-LMM model, which applies the pointwise LMM correction method proposed by Cui et al \cite{cui2022fast} to recover the latent EE curves. Previous studies \citep{chen2024adjusting,luan2023generalized} have shown that P-LMM yields more accurate estimates than the naive or average approaches when zero inflation is not present, making it a strong baseline for evaluating SC-based methods.

Figure~\ref{fig:SCcurves}(b) presents the mean corrected SC trajectories estimated using four methods. SC-Naive and SC-Average, which do not account for zero inflation, consistently underestimate activity levels compared to SC-P-ZIPMM and SC-BE-ZIME-3, resulting in biased step count trajectories. 
This difference arises because the naive and average methods treat all zero observations as true inactivity, without accounting for the possibility of nonwear periods or other structural zeros.
Among them, SC-Naive exhibits greater variability over time than SC-Average. SC-P-ZIPMM applies point-wise correction, yielding higher but more fluctuating estimates, while SC-BE-ZIME-3, based on basis expansion, produces a smoother trajectory. The smoothness of SC-BE-ZIME-3 is appropriate, as it reflects the underlying long-term activity pattern over time rather than moment-to-moment fluctuations. These comparisons highlight the importance of correcting for zero inflation to avoid systematic underestimation of the functional covariate.

To facilitate meaningful comparisons of functional coefficient estimates across correction methods, we scale all corrected SC curves using a common range, the minimum and maximum of the mean trajectory estimated by SC-BE-ZIME-3. Similarly, EE curves are scaled based on the range of the mean trajectory estimated by EE-P-LMM. Specifically, let $x_{\text{SC},L} = \min_t \bar{X}^*_{\text{SC}}(t)$ and $x_{\text{SC},U} = \max_t \bar{X}^*_{\text{SC}}(t)$, where $\bar{X}^*_{\text{SC}}(t) = \frac{1}{n} \sum_{i=1}^n \widehat{X}_{\text{SC-BE-ZIME-3},i}(t)$ denotes the mean SC trajectory from SC-BE-ZIME-3. Similarly, define $x_{\text{EE},L}$ and $x_{\text{EE},U}$ based on the mean trajectory of EE-P-LMM. Each curve is then scaled as
\begin{equation}\label{eq:scale}
    \widetilde{X}_{\text{SC},i}^{(m)}(t) = \frac{\widehat{X}^{(m)}_{\text{SC},i}(t) - x_{\text{SC},L}}{x_{\text{SC},U} - x_{\text{SC},L}}, 
    \qquad 
    \widetilde{X}_{\text{EE},i}(t) = \frac{\widehat{X}_{\text{EE},i}(t) - x_{\text{EE},L}}{x_{\text{EE},U} - x_{\text{EE},L}}.
\end{equation}
This ensures that all SC methods are scaled using the same range, enabling valid comparisons of their corresponding functional coefficients. Moreover, since both SC-BE-ZIME-3 and EE-P-LMM are scaled such that their mean trajectories lie within the same range $[0,1]$, their coefficient estimates are also directly comparable. In this way, SC-BE-ZIME-3 serves as a bridge, providing the reference range for scaling all SC methods while also aligning with EE-P-LMM through a shared range of mean trajectories. Notably, scaling SC curves based on the mean trajectory of any SC-based method does not substantially affect the comparison of the resulting functional coefficients, as will be discussed in Section~\ref{ssec:real_est}.

\subsection{Estimation of Joint Quantile Regression for BMI Change}\label{ssec:real_est}

We construct a joint linear quantile regression model with the outcome of interest being the change in BMI over 18 months. The functional covariate is the corrected and scaled EE or SC derived from one of the correction methods, accompanied by the error-free scalar covariates listed in Table~\ref{tab:real_cov}. The model is fit across three quantiles, \(\tau \in \{0.25, 0.50, 0.85\}\), to capture different regions of the BMI change distribution. Notably, the 0.85 quantile of age- and sex-adjusted BMI is commonly used as the clinical threshold for defining overweight status in children\citep{hampl2023clinical}. This section compares the estimated functional coefficients obtained under various SC- and EE-based correction methods.

\begin{figure}[h!]
    \centering
    \includegraphics[width=1\textwidth]{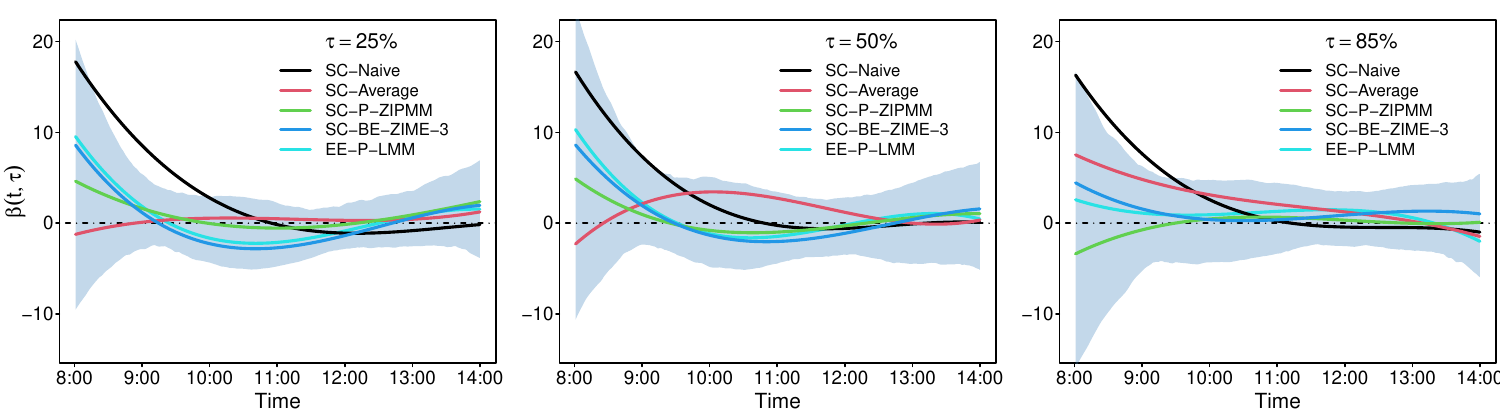}
    \caption{
    Estimated functional coefficients from different methods (solid lines), with 95\% pointwise bootstrap-based confidence intervals from the EE-P-LMM method (shaded areas), comparing SC-based methods to the EE-P-LMM method across different quantile levels \(\tau\). The horizontal line indicates zero. ``EE-Method'' and ``SC-Method'' refer to estimates based on corrected energy expenditure and step counts, respectively. The SC-Naive and SC-Average methods ignore both measurement error and zero inflation, while the SC-P-ZIPMM and SC-BE-ZIME-3 methods account for both.}   
    \label{fig:real_beta_main}
\end{figure}

Figure~\ref{fig:real_beta_main} illustrates the estimated functional coefficients \(\beta(t,\tau)\) from the SC-based methods in comparison with the EE-P-LMM method, along with 95\% pointwise bootstrap-based confidence intervals for EE-P-LMM. Results based on additional correction methods used in the simulation studies are provided in Supplementary Section~S2. 
Recall that SC-BE-ZIME-3 is used to bridge all methods through the scaling defined in Equation~\eqref{eq:scale}. The results show that the functional coefficients estimated by SC-BE-ZIME-3 are closely aligned with those from EE-P-LMM across all quantile levels. In contrast, SC-Naive and SC-Average display larger deviations from SC-BE-ZIME-3, as uncorrected, subject-specific and time-varying zero inflation leads to biased estimates of the functional covariates and, consequently, biased functional coefficients. Additional results using alternative scaling ranges based on other SC methods are provided in Supplementary Section S2, where similar patterns are observed.

To provide a direct measure of agreement between functional coefficient estimates, we compute the mean absolute pointwise deviation from the EE-P-LMM method:
\[
  \frac{1}{T} \sum_{t=1}^T \left| \hat{\beta}_{\cdot}(t,\tau) - \hat{\beta}_{\text{EE-P-LMM}}(t,\tau) \right|,
\]
where \(\hat{\beta}_{\cdot}(t,\tau)\) denotes the estimated functional coefficient at the \(\tau\)-th quantile under a given method (denoted by \(\cdot\)). This deviation measures the average pointwise difference between each method and the EE-P-LMM benchmark. 
The results show that SC-BE-ZIME-3 consistently produces estimates that are closer to EE-P-LMM than those from other SC-based methods. 
For instance, the average deviation across quantiles is 0.648 for SC-BE-ZIME-3, compared to 2.911 for SC-Naive and 2.206 for SC-Average. SC-P-ZIPMM also performs reasonably well, with an average deviation of 1.074. These findings further underscore the importance of correcting for both measurement error and zero inflation in the estimation process.

In Figure~\ref{fig:real_beta_main}, each pointwise bootstrap confidence interval for \(\hat{\beta}(t,\tau)\) includes zero, indicating a lack of statistical significance across the time points. 
However, we also conducted a wild bootstrap\citep{davidson2008wild}-based global test for \(H_0 : \beta(t) \equiv 0\), assessing the significance of the functional covariate effect at the mean level of the response distribution, using the following steps:
\begin{enumerate}
\item Fit the functional linear regression model with both scalar and SC- or EE-based functional covariates, yielding \(\hat{\beta}(t)\).
\item Fit the reduced linear model, omitting the functional covariate, to obtain fitted values \(\hat{y}_i\) and residuals \(\widehat{\varepsilon}_i = y_i - \hat{y}_i\).
\item Generate wild-bootstrap pseudo-responses \(y_i^{(b)} = \hat{y}_i + \xi_i^{(b)} \,\widehat{\varepsilon}_i\), where \(\xi_i^{(b)}\sim \text{Bernoulli}(0.5)\). Refit the linear regression including the functional covariate, record the estimate \(\hat{\beta}^{(b)}(t)\), for \(b=1,\dots,1000\).
\item Compute the \(p\)-value as 
\[
  \frac{1}{1000}\sum_{b=1}^{1000}
    \mathbf{I}\Bigl\{
      \|\hat{\beta}^{(b)}(t)\|^2
      \;>\;
      \|\hat{\beta}(t)\|^2
    \Bigr\},
\]
where \(\|\beta(t)\|^2 = \int (\beta(t))^2 \,dt\) is the \(L_2\)-norm.
\end{enumerate}
Table~\ref{tab:real_pv} reports the \(p\)-values of the functional coefficients for the various methods. For SC-BE-ZIME-3, SC-P-ZIPMM, and EE-P-LMM, their \(p\)-values are all below 0.05, indicating that SC and EE have a globally significant influence on BMI changes at the 5\% level. 
The discrepancy between global and pointwise significance occurs because the global test evaluates whether the entire coefficient function significantly deviates from zero in aggregate, while pointwise confidence intervals assess uncertainty at individual time points. The global test uses a statistic that accumulates deviations over the full domain, and the wild bootstrap approximates its null distribution through resampling. As a result, even if the pointwise intervals all contain zero, due to local variability or limited sample size, the global test can still detect small but consistent effects that are significant when considered over the whole function.

In conclusion, applying the proposed BE-ZIME framework to real-world data demonstrates its effectiveness in addressing biases arising from zero inflation and measurement error in SC data. The results further support the use of properly corrected SC as a reliable proxy for EE in modeling changes in BMI.

\section{Conclusion}\label{sec:conclusion}
To summarize, we propose a novel scalar-on-function quantile regression framework for analyzing functional covariates that are both zero-inflated and contaminated by measurement error—challenges commonly encountered in wearable and biomedical data. Our method introduces a subject-specific, time-varying validity indicator to distinguish structural zeros from genuine observations, and jointly estimates the latent functional covariates and zero-inflation probabilities through an iterative maximum likelihood procedure. Measurement error is addressed via linear mixed models applied to basis-expanded data. To characterize distributional effects, we incorporate joint quantile regression, ensuring coherent and stable estimation across quantile levels. Through extensive simulation studies, including both constant and piecewise constant zero-inflation probability settings, we demonstrate substantial improvements in estimation accuracy over competing methods. We further validate our framework using real data from a childhood obesity study, correcting for zero inflation and measurement error in step counts. Our results reveal strong agreement between corrected step counts and energy expenditure, demonstrating that properly corrected step counts can serve as a viable proxy for more complex activity measures like energy expenditure.

While our method provides a unified framework for addressing key challenges in functional data analysis, several avenues for future work remain. First, the current estimation of the zero-inflation probability is conducted in a pointwise or segmentwise manner, without explicitly modeling temporal correlation. Future research may explore structured priors or smoothing techniques to better capture the dynamics of time-varying zero inflation. Second, the number of basis functions is currently selected using classical methods that ignore the presence of measurement error and zero inflation. Developing more principled selection criteria that account for these complexities would further improve estimation accuracy. 




\bibliographystyle{unsrtnat}
\bibliography{reference}

\clearpage
{
\setlength{\tabcolsep}{3pt}
\begin{table}
\centering
\caption{{Estimation accuracy for the subject-specific functional covariate $X_i(t)$, evaluated under different zero-inflation levels, with sample size \(n=100\) and observed time points \(L=100\), based on 500 replications. In each replication \(b\), we compute the mean squared error 
$\text{MISE}_b(\widehat{X}) = (nL)^{-1} \sum_{i=1}^n\sum_{\ell = 1}^{L} \{ \widehat{X}^{(b)}_i(t_\ell) - {X}^{(b)}_i(t_\ell)\}^2$.
The table reports the average and standard deviation (in parentheses) of \(\{\text{MISE}_b(\widehat{X})\}_{b=1}^{500}\). Here, \(\pi_i\) represents the constant zero-inflation probability of the \(i\)-th subject, with \(\pi_i \sim \mathrm{Uniform}(\pi_0 - \pi_\delta,\, \pi_0 + \pi_\delta)\). Bolded values indicate the best performance (excluding the Oracle).}}\label{tab:s1xi}
\begin{tabular}{cr|rrrrrrr}
\toprule
$\pi_0$ & $\pi_\delta$ & Naive & Average & P-LMM & P-PMM & P-ZIPMM & BE-ME & BE-ZIME \\ \midrule
0 & 0 & 4.998 (0.126) & 0.714 (0.018) & 0.504 (0.014) & 0.514 (0.015) & 0.515 (0.015) & \textbf{0.255} (0.008) & 0.258 (0.008) \\
0.3 & 0 & 11.499 (0.395) & 3.698 (0.142) & 3.344 (0.123) & 3.424 (0.131) & 4.912 (0.170) & 2.903 (0.115) & \textbf{0.479} (0.031) \\
 & 0.1 & 11.512 (0.401) & 3.781 (0.149) & 3.347 (0.130) & 3.484 (0.138) & 4.913 (0.170) & 2.936 (0.125) & \textbf{0.486} (0.033) \\
 & 0.2 & 11.522 (0.410) & 4.011 (0.173) & 3.379 (0.146) & 3.657 (0.159) & 4.898 (0.167) & 3.087 (0.143) & \textbf{0.509} (0.037) \\
0.6 & 0 & 17.995 (0.657) & 10.798 (0.406) & 10.478 (0.378) & 10.633 (0.398) & 39.426 (1.362) & 9.798 (0.383) & \textbf{0.518} (0.185) \\
 & 0.1 & 18.003 (0.662) & 10.887 (0.418) & 10.468 (0.385) & 10.708 (0.409) & 39.215 (1.385) & 9.895 (0.387) & \textbf{0.518} (0.169) \\
 & 0.2 & 18.008 (0.666) & 11.118 (0.443) & 10.452 (0.397) & 10.910 (0.432) & 38.201 (1.373) & 10.159 (0.410) & \textbf{0.535} (0.156) \\ \bottomrule
\end{tabular}
\end{table}

\begin{table}
\centering
\caption{Estimation accuracy for the subject-specific zero-inflation probabilities \(\pi_i\), evaluated under different zero-inflation levels and combinations of sample size \(n\) and observed time points \(L\), based on 500 replications. In each replication \(b\), we compute the mean squared error \(\text{MSE}_b(\hat{\pi}) = \frac{1}{n} \sum_{i=1}^n (\hat{\pi}_i^{(b)} - \pi_i)^2\). The table reports the average and standard deviation (in parentheses) of \(\{\text{MSE}_b(\hat{\pi})\}_{b=1}^{500}\). Here, \(\pi_i\) represents the constant zero-inflation probability of the \(i\)-th subject, with \(\pi_i \sim \mathrm{Uniform}(\pi_0 - \pi_\delta,\, \pi_0 + \pi_\delta)\).}\label{tab:s1pi}
\begin{tabular}{ccccc}
\toprule
$\pi_0$ & $\pi_\delta$ & $n=100,\ L=100$ & $n=500,\ L=100$ & $n=100,\ L=200$ \\
\midrule
0.0 & 0 & 0.000048 (0.000096) & 0.000048 (0.000097) & 0.000040 (0.000069) \\
0.3 & 0 & 0.000322 (0.000457) & 0.000321 (0.000454) & 0.000161 (0.000230) \\
0.3 & 0.1 & 0.000314 (0.000448) & 0.000316 (0.000452) & 0.000161 (0.000231) \\
0.3 & 0.2 & 0.000302 (0.000445) & 0.000304 (0.000447) & 0.000153 (0.000225) \\
0.6 & 0 & 0.000364 (0.000514) & 0.000365 (0.000514) & 0.000182 (0.000256) \\
0.6 & 0.1 & 0.000358 (0.000506) & 0.000358 (0.000508) & 0.000182 (0.000258) \\
0.6 & 0.2 & 0.000344 (0.000493) & 0.000344 (0.000492) & 0.000175 (0.000252) \\
\bottomrule
\end{tabular}
\end{table}

\begin{table}
\centering
\caption{Estimation accuracy for \(\beta(t)=0.5\sin(\pi t)\) at \(\tau=0.5\) under different zero-inflation levels, with sample size \(n=100\) and observed time points \(L=100\), over 500 replications. Here, \(\pi_i\) represents the zero-inflation probability of the \(i\)-th subject, with \(\pi_i \sim \mathrm{Uniform}(\pi_0 - \pi_\delta,\, \pi_0 + \pi_\delta)\). Bolded values indicate the best performance (excluding the Oracle).}
\label{tab:s1beta1n100}
\begin{tabular}{ccc|cccccccc}
\toprule
& $\pi_0$ & $\pi_\delta$ & Naive & Average & P-LMM & P-PMM & P-ZIPMM & BE-ME & BE-ZIME & Oracle \\ \midrule
ABias$^2$ & 0 & 0 & 0.0019 & \textbf{0.0001} & 0.0272 & 0.0263 & 0.0252 & \textbf{0.0001} & \textbf{0.0001} & 0.0000 \\
 & 0.3 & 0 & 0.0087 & 0.0172 & 0.9815 & 0.0794 & \textbf{0.0006} & 0.0227 & \textbf{0.0006} & 0.0000 \\
 &  & 0.1 & 0.0038 & 0.0039 & 0.5152 & 0.0349 & \textbf{0.0006} & 0.0058 & \textbf{0.0006} & 0.0000 \\
 &  & 0.2 & 0.0088 & 0.0038 & 0.1064 & 0.0038 & \textbf{0.0005} & 0.0033 & 0.0007 & 0.0000 \\
 & 0.6 & 0 & 0.0560 & 0.1972 & 16.6678 & 0.3683 & 0.0154 & 0.2886 & \textbf{0.0035} & 0.0000 \\
 &  & 0.1 & 0.0136 & 0.0382 & 4.6619 & 0.0942 & 0.0163 & 0.0563 & \textbf{0.0039} & 0.0000 \\
 &  & 0.2 & 0.0119 & 0.0086 & 0.4808 & 0.0121 & 0.0207 & 0.0091 & \textbf{0.0050} & 0.0000 \\ \hline
AVar & 0 & 0 & \textbf{0.0059} & \textbf{0.0059} & 0.0125 & 0.0122 & 0.0120 & 0.0061 & 0.0061 & 0.0062 \\
 & 0.3 & 0 & 0.0108 & 0.0120 & 0.0954 & 0.0203 & 0.0070 & 0.0134 & \textbf{0.0056} & 0.0064 \\
 &  & 0.1 & 0.0155 & 0.0268 & 0.1682 & 0.0443 & 0.0076 & 0.0294 & \textbf{0.0062} & 0.0069 \\
 &  & 0.2 & 0.0246 & 0.0504 & 0.2209 & 0.0758 & 0.0082 & 0.0538 & \textbf{0.0064} & 0.0069 \\
 & 0.6 & 0 & 0.0265 & 0.0352 & 1.4143 & 0.0512 & \textbf{0.0026} & 0.0473 & 0.0057 & 0.0063 \\
 &  & 0.1 & 0.0361 & 0.1006 & 2.0103 & 0.1395 & \textbf{0.0032} & 0.1231 & 0.0057 & 0.0069 \\
 &  & 0.2 & 0.0514 & 0.1649 & 1.3325 & 0.2102 & \textbf{0.0046} & 0.1829 & 0.0058 & 0.0070 \\ \hline
MISE & 0 & 0 & 0.0079 & \textbf{0.0060} & 0.0397 & 0.0385 & 0.0371 & 0.0061 & 0.0062 & 0.0062 \\
 & 0.3 & 0 & 0.0195 & 0.0293 & 1.0770 & 0.0997 & 0.0076 & 0.0361 & \textbf{0.0062} & 0.0064 \\
 &  & 0.1 & 0.0193 & 0.0307 & 0.6834 & 0.0793 & 0.0082 & 0.0352 & \textbf{0.0068} & 0.0069 \\
 &  & 0.2 & 0.0334 & 0.0542 & 0.3273 & 0.0795 & 0.0087 & 0.0571 & \textbf{0.0071} & 0.0069 \\
 & 0.6 & 0 & 0.0825 & 0.2324 & 18.0821 & 0.4195 & 0.0180 & 0.3359 & \textbf{0.0092} & 0.0064 \\
 &  & 0.1 & 0.0497 & 0.1388 & 6.6722 & 0.2336 & 0.0195 & 0.1794 & \textbf{0.0096} & 0.0069 \\
 &  & 0.2 & 0.0633 & 0.1735 & 1.8133 & 0.2224 & 0.0254 & 0.1920 & \textbf{0.0108} & 0.0070 \\\bottomrule
\end{tabular}
\end{table}

\begin{table}
\centering
\caption{Estimation accuracy for \(\beta(t)=0.5\sin(\pi t)\) at \(\tau=0.5\) under different zero-inflation levels, with sample size \(n=500\) and observed time points \(L=100\), over 500 replications. Here, \(\pi_i\) represents the zero-inflation probability of the \(i\)-th subject, with \(\pi_i \sim \mathrm{Uniform}(\pi_0 - \pi_\delta,\, \pi_0 + \pi_\delta)\). Bolded values indicate the best performance (excluding the Oracle).}
\label{tab:s1beta1n500}
\begin{tabular}{ccc|cccccccc}
\toprule
& $\pi_0$ & $\pi_\delta$ & Naive & Average & P-LMM & P-PMM & P-ZIPMM & BE-ME & BE-ZIME & Oracle \\ \midrule
ABias$^2$ & 0 & 0 & 0.0022 & \textbf{0.0001} & 0.0263 & 0.0249 & 0.0239 & \textbf{0.0001} & \textbf{0.0001} & 0.0000 \\
 & 0.3 & 0 & 0.0086 & 0.0172 & 0.9649 & 0.0780 & \textbf{0.0005} & 0.0226 & 0.0006 & 0.0000 \\
 &  & 0.1 & 0.0036 & 0.0040 & 0.5181 & 0.0349 & \textbf{0.0005} & 0.0060 & 0.0006 & 0.0000 \\
 &  & 0.2 & 0.0089 & 0.0044 & 0.1137 & 0.0050 & \textbf{0.0004} & 0.0039 & 0.0007 & 0.0000 \\
 & 0.6 & 0 & 0.0533 & 0.1951 & 18.7388 & 0.3623 & 0.0157 & 0.2856 & \textbf{0.0037} & 0.0000 \\
 &  & 0.1 & 0.0134 & 0.0377 & 4.7294 & 0.0918 & 0.0165 & 0.0552 & \textbf{0.0039} & 0.0000 \\
 &  & 0.2 & 0.0123 & 0.0079 & 0.4539 & 0.0102 & 0.0199 & 0.0082 & \textbf{0.0048} & 0.0000 \\ \hline
AVar & 0 & 0 & \textbf{0.0011} & \textbf{0.0011} & 0.0024 & 0.0023 & 0.0023 & 0.0012 & 0.0012 & 0.0012 \\
 & 0.3 & 0 & 0.0021 & 0.0024 & 0.0187 & 0.0042 & 0.0013 & 0.0027 & \textbf{0.0011} & 0.0012 \\
 &  & 0.1 & 0.0029 & 0.0054 & 0.0335 & 0.0088 & 0.0015 & 0.0059 & \textbf{0.0012} & 0.0013 \\
 &  & 0.2 & 0.0047 & 0.0102 & 0.0445 & 0.0150 & 0.0016 & 0.0109 & \textbf{0.0011} & 0.0012 \\
 & 0.6 & 0 & 0.0053 & 0.0066 & 0.3392 & 0.0096 & \textbf{0.0005} & 0.0088 & 0.0010 & 0.0012 \\
 &  & 0.1 & 0.0075 & 0.0198 & 0.4126 & 0.0274 & \textbf{0.0006} & 0.0241 & 0.0010 & 0.0012 \\
 &  & 0.2 & 0.0093 & 0.0288 & 0.2364 & 0.0356 & \textbf{0.0008} & 0.0320 & 0.0010 & 0.0012 \\ \hline
MISE & 0 & 0 & 0.0033 & \textbf{0.0012} & 0.0287 & 0.0273 & 0.0261 & \textbf{0.0012} & 0.0013 & 0.0012 \\
 & 0.3 & 0 & 0.0107 & 0.0196 & 0.9836 & 0.0822 & 0.0018 & 0.0253 & \textbf{0.0017} & 0.0012 \\
 &  & 0.1 & 0.0065 & 0.0094 & 0.5516 & 0.0437 & 0.0019 & 0.0119 & \textbf{0.0018} & 0.0013 \\
 &  & 0.2 & 0.0136 & 0.0145 & 0.1583 & 0.0200 & 0.0020 & 0.0148 & \textbf{0.0018} & 0.0013 \\
 & 0.6 & 0 & 0.0586 & 0.2018 & 19.0780 & 0.3719 & 0.0162 & 0.2944 & \textbf{0.0048} & 0.0012 \\
 &  & 0.1 & 0.0209 & 0.0575 & 5.1420 & 0.1192 & 0.0171 & 0.0793 & \textbf{0.0049} & 0.0012 \\
 &  & 0.2 & 0.0216 & 0.0367 & 0.6903 & 0.0459 & 0.0207 & 0.0402 & \textbf{0.0059} & 0.0012 \\
\bottomrule
\end{tabular}
\end{table}

\begin{table}
\centering
\caption{Estimation accuracy for \(\beta(t)=\sin(2\pi t)\) at \(\tau=0.5\) under different zero-inflation levels, with sample size \(n=100\) and observed time points \(L=100\), over 500 replications. Here, \(\pi_i\) represents the zero-inflation probability of the \(i\)-th subject, with \(\pi_i \sim \mathrm{Uniform}(\pi_0 - \pi_\delta,\, \pi_0 + \pi_\delta)\). Bolded values indicate the best performance (excluding the Oracle).}
\label{tab:s1beta2gp100}
\begin{tabular}{ccc|cccccccc}
\toprule
& $\pi_0$ & $\pi_\delta$ & Naive & Average & P-LMM & P-PMM & P-ZIPMM & BE-ME & BE-ZIME & Oracle \\ \midrule
ABias$^2$ & 0 & 0 & 0.0352 & 0.0055 & 0.0823 & 0.0788 & 0.0752 & \textbf{0.0051} & 0.0052 & 0.0051 \\
 & 0.3 & 0 & 0.0502 & 0.0376 & 2.6786 & 0.2002 & \textbf{0.0048} & 0.0520 & 0.0119 & 0.0050 \\
 &  & 0.1 & 0.0512 & 0.0347 & 2.1538 & 0.1799 & \textbf{0.0048} & 0.0464 & 0.0117 & 0.0052 \\
 &  & 0.2 & 0.0507 & 0.0302 & 1.2327 & 0.1327 & \textbf{0.0048} & 0.0374 & 0.0118 & 0.0052 \\
 & 0.6 & 0 & 0.0623 & 0.3566 & 32.8742 & 0.7314 & 0.0900 & 0.5749 & \textbf{0.0327} & 0.0050 \\
 &  & 0.1 & 0.0625 & 0.3265 & 20.7700 & 0.6431 & 0.0904 & 0.4607 & \textbf{0.0329} & 0.0052 \\
 &  & 0.2 & 0.0614 & 0.2697 & 7.5195 & 0.4582 & 0.0965 & 0.3252 & \textbf{0.0357} & 0.0052 \\ \hline
AVar & 0 & 0 & 0.0107 & \textbf{0.0075} & 0.0166 & 0.0173 & 0.0170 & 0.0077 & 0.0076 & 0.0061 \\
 & 0.3 & 0 & 0.0206 & 0.0190 & 0.1637 & 0.0330 & \textbf{0.0093} & 0.0211 & 0.0101 & 0.0063 \\
 &  & 0.1 & 0.0215 & 0.0193 & 0.1563 & 0.0333 & \textbf{0.0100} & 0.0212 & 0.0112 & 0.0070 \\
 &  & 0.2 & 0.0205 & 0.0199 & 0.1107 & 0.0318 & \textbf{0.0103} & 0.0213 & 0.0114 & 0.0071 \\
 & 0.6 & 0 & 0.0467 & 0.0672 & 3.0959 & 0.0977 & \textbf{0.0045} & 0.0907 & 0.0151 & 0.0063 \\
 &  & 0.1 & 0.0459 & 0.0692 & 1.9304 & 0.0974 & \textbf{0.0044} & 0.0855 & 0.0154 & 0.0071 \\
 &  & 0.2 & 0.0439 & 0.0661 & 0.7027 & 0.0884 & \textbf{0.0049} & 0.0735 & 0.0170 & 0.0071 \\ \hline
MISE & 0 & 0 & 0.0459 & 0.0130 & 0.0989 & 0.0961 & 0.0922 & \textbf{0.0128} & 0.0129 & 0.0111 \\
 & 0.3 & 0 & 0.0707 & 0.0565 & 2.8422 & 0.2331 & \textbf{0.0141} & 0.0731 & 0.0220 & 0.0114 \\
 &  & 0.1 & 0.0728 & 0.0540 & 2.3101 & 0.2132 & \textbf{0.0148} & 0.0676 & 0.0229 & 0.0122 \\
 &  & 0.2 & 0.0711 & 0.0501 & 1.3435 & 0.1645 & \textbf{0.0151} & 0.0587 & 0.0231 & 0.0122 \\
 & 0.6 & 0 & 0.1090 & 0.4238 & 35.9700 & 0.8290 & 0.0945 & 0.6656 & \textbf{0.0478} & 0.0113 \\
 &  & 0.1 & 0.1084 & 0.3957 & 22.7004 & 0.7405 & 0.0949 & 0.5462 & \textbf{0.0483} & 0.0122 \\
 &  & 0.2 & 0.1053 & 0.3358 & 8.2222 & 0.5467 & 0.1014 & 0.3987 & \textbf{0.0528} & 0.0122 \\
\bottomrule
\end{tabular}
\end{table}

\begin{table}
\centering
\caption{Estimation accuracy for \(\beta(t)=\sin(2\pi t)\) at \(\tau=0.5\) under different zero-inflation levels, with sample size \(n=100\) and observed time points \(L=200\), over 500 replications. Here, \(\pi_i\) represents the zero-inflation probability of the \(i\)-th subject, with \(\pi_i \sim \mathrm{Uniform}(\pi_0 - \pi_\delta,\, \pi_0 + \pi_\delta)\). Bolded values indicate the best performance (excluding the Oracle).}
\label{tab:s1beta2gp200}
\begin{tabular}{ccc|cccccccc}
\toprule
& $\pi_0$ & $\pi_\delta$ & Naive & Average & P-LMM & P-PMM & P-ZIPMM & BE-ME & BE-ZIME & Oracle \\ \midrule
ABias$^2$ & 0 & 0 & 0.0155 & 0.0057 & 0.0949 & 0.0918 & 0.0878 & \textbf{0.0056} & 0.0057 & 0.0059 \\
 & 0.3 & 0 & 0.0234 & 0.0579 & 3.2503 & 0.2698 & \textbf{0.0060} & 0.0670 & 0.0073 & 0.0058 \\
 &  & 0.1 & 0.0211 & 0.0537 & 2.5635 & 0.2377 & \textbf{0.0060} & 0.0610 & 0.0078 & 0.0059 \\
 &  & 0.2 & 0.0203 & 0.0461 & 1.4640 & 0.1770 & \textbf{0.0060} & 0.0507 & 0.0080 & 0.0059 \\
 & 0.6 & 0 & 0.0481 & 0.5619 & 47.1640 & 1.1023 & 0.0754 & 0.6989 & \textbf{0.0192} & 0.0058 \\
 &  & 0.1 & 0.0473 & 0.5440 & 29.3346 & 1.0011 & 0.0752 & 0.6320 & \textbf{0.0180} & 0.0059 \\
 &  & 0.2 & 0.0433 & 0.4469 & 10.3983 & 0.7239 & 0.0794 & 0.4839 & \textbf{0.0194} & 0.0059 \\ \hline
AVar & 0 & 0 & 0.0104 & \textbf{0.0072} & 0.0155 & 0.0160 & 0.0157 & 0.0073 & \textbf{0.0072} & 0.0062 \\
 & 0.3 & 0 & 0.0227 & 0.0160 & 0.1484 & 0.0290 & 0.0093 & 0.0169 & \textbf{0.0083} & 0.0064 \\
 &  & 0.1 & 0.0202 & 0.0164 & 0.1338 & 0.0288 & 0.0093 & 0.0172 & \textbf{0.0082} & 0.0062 \\
 &  & 0.2 & 0.0217 & 0.0177 & 0.1024 & 0.0287 & 0.0097 & 0.0183 & \textbf{0.0089} & 0.0062 \\
 & 0.6 & 0 & 0.0626 & 0.0612 & 2.6322 & 0.0898 & \textbf{0.0040} & 0.0711 & 0.0114 & 0.0063 \\
 &  & 0.1 & 0.0610 & 0.0622 & 1.7065 & 0.0884 & \textbf{0.0040} & 0.0691 & 0.0118 & 0.0062 \\
 &  & 0.2 & 0.0529 & 0.0667 & 0.7284 & 0.0894 & \textbf{0.0043} & 0.0702 & 0.0118 & 0.0062 \\ \hline
MISE & 0 & 0 & 0.0259 & \textbf{0.0129} & 0.1104 & 0.1078 & 0.1035 & \textbf{0.0129} & \textbf{0.0129} & 0.0121 \\
 & 0.3 & 0 & 0.0461 & 0.0739 & 3.3986 & 0.2988 & \textbf{0.0152} & 0.0839 & 0.0156 & 0.0122 \\
 &  & 0.1 & 0.0413 & 0.0701 & 2.6973 & 0.2665 & \textbf{0.0153} & 0.0782 & 0.0161 & 0.0121 \\
 &  & 0.2 & 0.0420 & 0.0639 & 1.5664 & 0.2057 & \textbf{0.0157} & 0.0690 & 0.0169 & 0.0121 \\
 & 0.6 & 0 & 0.1107 & 0.6231 & 49.7962 & 1.1921 & 0.0794 & 0.7700 & \textbf{0.0305} & 0.0121 \\
 &  & 0.1 & 0.1083 & 0.6062 & 31.0410 & 1.0896 & 0.0793 & 0.7011 & \textbf{0.0298} & 0.0121 \\
 &  & 0.2 & 0.0962 & 0.5136 & 11.1267 & 0.8134 & 0.0837 & 0.5541 & \textbf{0.0312} & 0.0121 \\
\bottomrule
\end{tabular}
\end{table}

\begin{table}
\centering
\caption{Relative improvement (\%) from joint estimation for \(\beta(t)=0.5\sin(\pi t)\) at \(\tau \in \{0.25, 0.75\}\) under different zero-inflation levels, with sample size \(n=100\) and observed time points \(L=100\), over 500 replications. Here, \(\pi_i\) denotes the zero-inflation probability of the \(i\)-th subject, with \(\pi_i \sim \mathrm{Uniform}(\pi_0 - \pi_\delta,\, \pi_0 + \pi_\delta)\).}
\label{tab:s1jsbeta1}
\begin{tabular}{ccc|cccccccc}
\toprule
$\pi_0$ & $\pi_\delta$ & $\tau$ & Naive & Average & P-LMM & P-PMM & P-ZIPMM & BE-ME & BE-ZIME & Oracle \\ \midrule
0 & 0 & 0.25 & 10.03 & 19.89 & 5.36 & 7.00 & 7.24 & 19.84 & 19.02 & 17.24 \\
 &  & 0.75 & 11.20 & 21.95 & 8.01 & 9.09 & 8.81 & 22.08 & 22.36 & 18.81 \\ \hline
0.3 & 0 & 0.25 & 2.55 & 6.15 & 0.57 & 3.46 & 20.10 & 5.42 & 16.48 & 18.86 \\
 &  & 0.75 & 13.01 & 11.59 & 2.72 & 6.60 & 18.77 & 10.68 & 21.09 & 16.79 \\ 
 & 0.1 & 0.25 & 8.12 & 6.71 & 1.97 & 3.30 & 12.00 & 6.36 & 11.70 & 17.68 \\
 &  & 0.75 & 18.69 & 21.05 & 7.37 & 16.21 & 20.00 & 20.37 & 20.49 & 17.18 \\ 
 & 0.2 & 0.25 & 5.79 & 1.36 & 1.97 & -1.01 & 13.66 & 1.31 & 10.24 & 17.47 \\
 &  & 0.75 & 14.50 & 33.02 & 28.75 & 34.62 & 22.32 & 33.41 & 18.23 & 17.14 \\ \hline
0.6 & 0 & 0.25 & -1.26 & 1.13 & -0.04 & 0.44 & 3.56 & 0.80 & 6.89 & 19.26 \\
 &  & 0.75 & 11.02 & 4.45 & 2.24 & 3.71 & 4.22 & 4.15 & 13.46 & 15.13 \\ 
 & 0.1 & 0.25 & 9.53 & 4.50 & 2.48 & 2.79 & 3.11 & 4.39 & 8.12 & 17.83 \\
 &  & 0.75 & 13.56 & 20.95 & 11.11 & 17.40 & 3.79 & 20.19 & 14.88 & 16.78 \\ 
 & 0.2 & 0.25 & 8.17 & -1.24 & 1.86 & -1.75 & 1.32 & -1.24 & 3.96 & 17.17 \\
 &  & 0.75 & 5.81 & 17.37 & 16.49 & 18.89 & 6.71 & 17.50 & 11.32 & 16.89 \\
\bottomrule
\end{tabular}
\end{table}
\begin{table}
\centering
\caption{Relative improvement (\%) from joint estimation for \(\beta(t)=\sin(2\pi t)\) at \(\tau \in \{0.25, 0.75\}\) under different zero-inflation levels, with sample size \(n=100\) and observed time points \(L=100\), over 500 replications. Here, \(\pi_i\) denotes the zero-inflation probability of the \(i\)-th subject, with \(\pi_i \sim \mathrm{Uniform}(\pi_0 - \pi_\delta,\, \pi_0 + \pi_\delta)\).}
\label{tab:s1jsbeta2}
\begin{tabular}{ccc|cccccccc}
\toprule
$\pi_0$ & $\pi_\delta$ & $\tau$ & Naive & Average & P-LMM & P-PMM & P-ZIPMM & BE-ME & BE-ZIME & Oracle \\ \midrule
0 & 0 & 0.25 & 4.11 & 8.51 & 1.96 & 0.89 & 1.42 & 8.71 & 9.53 & 9.61 \\
 &  & 0.75 & 3.94 & 10.99 & 3.71 & 3.93 & 4.27 & 11.46 & 11.73 & 13.78 \\ \hline
0.3 & 0 & 0.25 & 7.51 & 7.09 & 0.09 & 2.93 & 13.20 & 6.06 & 11.11 & 10.06 \\
 &  & 0.75 & 4.93 & 7.27 & 1.30 & 2.99 & 14.40 & 6.22 & 12.62 & 12.04 \\ 
 & 0.1 & 0.25 & 8.52 & 10.62 & 0.98 & 5.11 & 13.49 & 9.45 & 9.05 & 11.12 \\
 &  & 0.75 & 4.57 & 8.92 & 0.84 & 3.14 & 12.72 & 7.83 & 8.62 & 8.83 \\ 
 & 0.2 & 0.25 & 8.85 & 10.41 & 1.40 & 4.61 & 14.08 & 9.63 & 5.95 & 10.73 \\
 &  & 0.75 & 4.62 & 10.42 & 1.91 & 5.05 & 12.77 & 9.59 & 8.03 & 8.48 \\ \hline
0.6 & 0 & 0.25 & 7.35 & 4.40 & 0.44 & 2.74 & 1.15 & 3.76 & 5.68 & 10.43 \\
 &  & 0.75 & 5.65 & 4.39 & 1.29 & 3.84 & 0.46 & 3.71 & 4.97 & 11.27 \\ 
 & 0.1 & 0.25 & 8.00 & 2.81 & 0.92 & 1.64 & 0.62 & 2.55 & 5.12 & 10.96 \\
 &  & 0.75 & 5.92 & 2.80 & 0.52 & 2.06 & 0.69 & 2.36 & 6.42 & 9.10 \\ 
 & 0.2 & 0.25 & 10.36 & 4.81 & 1.59 & 4.32 & 0.85 & 4.56 & 5.72 & 11.13 \\
 &  & 0.75 & 10.29 & 5.11 & 1.48 & 3.67 & 0.90 & 4.82 & 3.97 & 9.28 \\
\bottomrule
\end{tabular}
\end{table}
}

{
\setlength{\tabcolsep}{2pt}
\begin{table}
\centering
{
\caption{{Estimation accuracy for the subject-specific functional covariate $X_i(t)$ with sample sizes $n\in\{100,500\}$ and $L=100$ observed time points, based on 500 replications under piecewise-constant zero inflation schemes. 
Specifically, Case 1 (2 equal segments) $\pi_i(t)=\pi_0$ on $[0,0.5)$ and $1-\pi_0$ on $[0.5,1]$; Case 2 (5 equal segments) $\pi_i(t)=\pi_0$ on $[0,0.2)\cup[0.4,0.6)\cup[0.8,1]$ and $1-\pi_0$ on $[0.2,0.4)\cup[0.6,0.8)$; Case 3 (3 unequal segments) $\pi_i(t)=\pi_0$ on $[0,0.1)\cup[0.6,1]$ and $1-\pi_0$ on $[0.1,0.6)$.
In each replication \(b\), we compute the mean squared error 
$\text{MISE}_b(\widehat{X}) = (nL)^{-1} \sum_{i=1}^n\sum_{\ell = 1}^{L} \{ \widehat{X}^{(b)}_i(t_\ell) - {X}^{(b)}_i(t_\ell)\}^2$.
The table reports the average and standard deviation (in parentheses) of \(\{\text{MISE}_b(\widehat{X})\}_{b=1}^{500}\). 
Here $\widehat{M}$ denotes the working number of segments assumed by BE-ZIME during estimation. Bolded values denote the best performance (excluding the Oracle).
}}\label{tab:s2xi}
\resizebox{\textwidth}{!}{
\begin{tabular}{ccc|cccccccc}
\toprule
$n$ & $\pi_0$ & Case & Naive & Average & P-ZIPMM & BE-ME &
$\widehat{M}=1$ & $\widehat{M}=2$ & $\widehat{M}=5$ & $\widehat{M}=10$  \\ \midrule
100 & 0.6 & 1 & 15.830 (0.584) & 8.200 (0.310) & 24.829 (0.878) & 7.199 (0.325) & 1.470 (0.221) & 0.380 (0.057) & 0.336 (0.056) & \textbf{0.294} (0.009) \\
 &  & 2 & 16.269 (0.586) & 8.731 (0.333) & 27.828 (0.982) & 7.495 (0.302) & 0.655 (0.130) & 0.588 (0.033) & 0.332 (0.030) & \textbf{0.304} (0.042) \\
 &  & 3 & 15.846 (0.560) & 8.218 (0.319) & 24.909 (0.918) & 7.128 (0.288) & 1.285 (0.124) & 0.762 (0.071) & 0.386 (0.067) & \textbf{0.300} (0.010) \\
 & 0.8 & 1 & 15.834 (0.574) & 10.029 (0.386) & 53.684 (2.373) & 8.985 (0.366) & 7.247 (0.325) & 0.494 (0.033) & 0.479 (0.024) & \textbf{0.347} (0.013) \\
 &  & 2 & 17.160 (0.624) & 11.616 (0.442) & 63.980 (2.693) & 9.035 (0.351) & 2.672 (0.123) & 2.607 (0.121) & 0.459 (0.022) & \textbf{0.384} (0.015) \\
 &  & 3 & 15.867 (0.573) & 10.069 (0.401) & 53.610 (2.461) & 8.871 (0.360) & 7.041 (0.307) & 3.738 (0.164) & 0.877 (0.041) & \textbf{0.377} (0.015) \\ \hline
500 & 0.6 & 1 & 15.826 (0.254) & 8.198 (0.139) & 24.659 (0.398) & 7.181 (0.222) & 1.456 (0.214) & 0.374 (0.024) & 0.334 (0.048) & \textbf{0.294} (0.004) \\
 &  & 2 & 16.274 (0.262) & 8.729 (0.146) & 27.667 (0.426) & 7.486 (0.143) & 0.644 (0.113) & 0.586 (0.014) & 0.332 (0.029) & \textbf{0.301} (0.004) \\
 &  & 3 & 15.836 (0.259) & 8.213 (0.141) & 24.777 (0.402) & 7.116 (0.136) & 1.273 (0.081) & 0.760 (0.055) & 0.382 (0.040) & \textbf{0.300} (0.004) \\
 & 0.8 & 1 & 15.824 (0.259) & 10.023 (0.182) & 53.238 (1.076) & 8.994 (0.205) & 7.237 (0.179) & 0.494 (0.014) & 0.479 (0.010) & \textbf{0.347} (0.006) \\
 &  & 2 & 17.156 (0.278) & 11.612 (0.197) & 63.572 (1.183) & 9.031 (0.155) & 2.667 (0.077) & 2.601 (0.077) & 0.458 (0.011) & \textbf{0.384} (0.007) \\
 &  & 3 & 15.860 (0.260) & 10.062 (0.178) & 53.382 (1.078) & 8.864 (0.159) & 7.033 (0.139) & 3.733 (0.072) & 0.875 (0.020) & \textbf{0.377} (0.007) \\ \bottomrule
\end{tabular}
}
}
\end{table}

\begin{table}
\centering
{
\caption{{Estimation accuracy for zero-inflation probabilities \(\pi_i(t)\) with sample sizes $n\in\{100,500\}$ and $L=100$ observed time points, based on 500 replications under piecewise-constant zero inflation schemes.
Specifically, Case 1 (2 equal segments) $\pi_i(t)=\pi_0$ on $[0,0.5)$ and $1-\pi_0$ on $[0.5,1]$; Case 2 (5 equal segments) $\pi_i(t)=\pi_0$ on $[0,0.2)\cup[0.4,0.6)\cup[0.8,1]$ and $1-\pi_0$ on $[0.2,0.4)\cup[0.6,0.8)$; Case 3 (3 unequal segments) $\pi_i(t)=\pi_0$ on $[0,0.1)\cup[0.6,1]$ and $1-\pi_0$ on $[0.1,0.6)$.
For each \(\widehat{M}\), we define the corresponding equal-length partition \(\{\widetilde{\mathcal{T}}_m\}_{m=1}^{\widehat{M}}\) and compute \(\tilde{\pi}_{im} = \int_{\widetilde{\mathcal{T}}_m} \pi_i(t) dt\), setting \(\tilde{\pi}_i(t) = \tilde{\pi}_{im}\) for \(t \in \widetilde{\mathcal{T}}_m\). Each cell reports the average and standard deviation (in parentheses) of either \(\text{MSE}_b(\hat{\pi}) = \int (\hat{\pi}_i(t) - \pi_i(t))^2 dt\) or \(\text{MSE}_b^\dagger(\hat{\pi}) = \int (\hat{\pi}_i(t) - \tilde{\pi}_i(t))^2 dt\) across 500 replications.
Here $\widehat{M}$ denotes the working number of segments assumed by BE-ZIME during estimation.
Bolded values denote the smallest \(\text{MSE}_b(\hat{\pi})\) for each setting.
}}\label{tab:s2pi}
\resizebox{\textwidth}{!}{
\begin{tabular}{@{}ccc|cccc|cccc@{}}
\toprule
\multicolumn{3}{c|}{} 
& \multicolumn{4}{c|}{\(\text{MSE}_b(\hat{\pi})\)} 
& \multicolumn{4}{c}{\(\text{MSE}_b^\dagger(\hat{\pi})\)} \\
\cmidrule{4-11}
$n$ & $\pi_0$ & Case 
& $\widehat{M}=1$ & $\widehat{M}=2$ & $\widehat{M}=5$ & $\widehat{M}=10$ 
& $\widehat{M}=1$ & $\widehat{M}=2$ & $\widehat{M}=5$ & $\widehat{M}=10$ \\
\midrule
100 & 0.6 & 1 & 1.0e-02 (1e-02) & \textbf{6.6e-04} (7e-04) & 3.7e-03 (4e-03) & 3.4e-03 (3e-03) & 3.9e-04 (4e-04) & 6.6e-04 (7e-04) & 1.7e-03 (2e-03) & 3.4e-03 (3e-03) \\
 &  & 2 & 1.0e-02 (1e-02) & 1.0e-02 (1e-02) & \textbf{1.7e-03} (2e-03) & 3.4e-03 (3e-03) & 3.8e-04 (4e-04) & 5.3e-04 (5e-04) & 1.7e-03 (2e-03) & 3.4e-03 (3e-03) \\
 &  & 3 & 1.0e-02 (1e-02) & 7.1e-03 (7e-03) & 3.7e-03 (4e-03) & \textbf{3.4e-03} (3e-03) & 3.7e-04 (4e-04) & 6.7e-04 (7e-04) & 1.7e-03 (2e-03) & 3.4e-03 (3e-03) \\
 & 0.8 & 1 & 9.5e-02 (9e-02) & \textbf{4.7e-04} (5e-04) & 1.9e-02 (2e-02) & 2.3e-03 (2e-03) & 4.7e-03 (5e-03) & 4.7e-04 (5e-04) & 1.2e-03 (1e-03) & 2.3e-03 (2e-03) \\
 &  & 2 & 8.7e-02 (9e-02) & 8.7e-02 (9e-02) & \textbf{1.2e-03} (1e-03) & 2.3e-03 (2e-03) & 3.1e-04 (3e-04) & 5.1e-04 (5e-04) & 1.2e-03 (1e-03) & 2.3e-03 (2e-03) \\
 &  & 3 & 9.4e-02 (9e-02) & 6.0e-02 (6e-02) & 1.9e-02 (2e-02) & \textbf{2.4e-03} (2e-03) & 4.1e-03 (4e-03) & 2.0e-03 (2e-03) & 1.5e-03 (1e-03) & 2.4e-03 (2e-03) \\ \hline
500 & 0.6 & 1 & 1.0e-02 (1e-02) & \textbf{6.6e-04} (7e-04) & 3.7e-03 (4e-03) & 3.4e-03 (3e-03) & 3.9e-04 (4e-04) & 6.6e-04 (7e-04) & 1.7e-03 (2e-03) & 3.4e-03 (3e-03) \\
 &  & 2 & 1.0e-02 (1e-02) & 1.0e-02 (1e-02) & \textbf{1.7e-03} (2e-03) & 3.4e-03 (3e-03) & 3.8e-04 (4e-04) & 5.3e-04 (5e-04) & 1.7e-03 (2e-03) & 3.4e-03 (3e-03) \\
 &  & 3 & 1.0e-02 (1e-02) & 7.1e-03 (7e-03) & 3.7e-03 (4e-03) & \textbf{3.4e-03} (3e-03) & 3.7e-04 (4e-04) & 6.7e-04 (7e-04) & 1.7e-03 (2e-03) & 3.4e-03 (3e-03) \\
 & 0.8 & 1 & 9.5e-02 (9e-02) & \textbf{4.7e-04} (5e-04) & 1.9e-02 (2e-02) & 2.3e-03 (2e-03) & 4.7e-03 (5e-03) & 4.7e-04 (5e-04) & 1.2e-03 (1e-03) & 2.3e-03 (2e-03) \\
 &  & 2 & 8.7e-02 (9e-02) & 8.7e-02 (9e-02) & \textbf{1.2e-03} (1e-03) & 2.3e-03 (2e-03) & 3.1e-04 (3e-04) & 5.1e-04 (5e-04) & 1.2e-03 (1e-03) & 2.3e-03 (2e-03) \\
 &  & 3 & 9.4e-02 (9e-02) & 6.0e-02 (6e-02) & 1.9e-02 (2e-02) & \textbf{2.3e-03} (2e-03) & 4.1e-03 (4e-03) & 2.0e-03 (2e-03) & 1.5e-03 (1e-03) & 2.3e-03 (2e-03) \\ \bottomrule
\end{tabular}}}
\end{table}
}

{
\setlength{\tabcolsep}{3pt}
\begin{table}
\centering
{
\caption{{
Estimation accuracy for $\beta(t,\tau)$ at $\tau=0.5$ in the homogeneous setting $(\eta_\beta,\eta_\varepsilon)=(0,0)$, with sample sizes $n\in\{100,500\}$ and $L=100$ observed time points, based on 500 replications under piecewise-constant zero inflation schemes. 
Specifically, Case 1 (2 equal segments) $\pi_i(t)=\pi_0$ on $[0,0.5)$ and $1-\pi_0$ on $[0.5,1]$; Case 2 (5 equal segments) $\pi_i(t)=\pi_0$ on $[0,0.2)\cup[0.4,0.6)\cup[0.8,1]$ and $1-\pi_0$ on $[0.2,0.4)\cup[0.6,0.8)$; Case 3 (3 unequal segments) $\pi_i(t)=\pi_0$ on $[0,0.1)\cup[0.6,1]$ and $1-\pi_0$ on $[0.1,0.6)$.
Here $\widehat{M}$ denotes the working number of segments assumed by BE-ZIME during estimation. Bolded values denote the best performance (excluding the Oracle).
}}
\label{tab:s2beta1eta00}
\begin{tabular}{cccc|ccccccccc}
\toprule
& $n$ & $\pi_0$ & Case & Naive & Average & P-ZIPMM & BE-ME &
$\widehat{M}=1$ & $\widehat{M}=2$ & $\widehat{M}=5$ & $\widehat{M}=10$ &  Oracle \\ \midrule
ABias$^2$ & 100 & 0.6 & 1 & 0.0300 & 0.0962 & 0.0089 & 0.1246 & 0.0034 & 0.0027 & 0.0009 & \textbf{0.0006} & 0.0001 \\
 &  &  & 2 & 0.0303 & 0.0869 & 0.0074 & 0.0883 & 0.0047 & 0.0031 & 0.0008 & \textbf{0.0005} & 0.0001 \\
 &  &  & 3 & 0.0273 & 0.0786 & 0.0055 & 0.0788 & 0.0047 & 0.0038 & 0.0007 & \textbf{0.0004} & 0.0001 \\
 &  & 0.8 & 1 & 0.0279 & 0.2378 & 0.0346 & 0.2380 & \textbf{0.0012} & 0.0061 & 0.0025 & 0.0024 & 0.0001 \\
 &  &  & 2 & 0.0369 & 0.0942 & 0.0285 & 0.0942 & 0.0037 & 0.0035 & 0.0015 & \textbf{0.0007} & 0.0001 \\
 &  &  & 3 & 0.0195 & 0.1321 & 0.0220 & 0.1321 & 0.0062 & 0.0067 & 0.0037 & \textbf{0.0012} & 0.0001 \\ \cmidrule{2-13}
 & 500 & 0.6 & 1 & 0.0306 & 0.0953 & 0.0087 & 0.0954 & 0.0037 & 0.0028 & 0.0008 & \textbf{0.0006} & 0.0001 \\
 &  &  & 2 & 0.0318 & 0.0873 & 0.0073 & 0.0871 & 0.0036 & 0.0032 & 0.0008 & \textbf{0.0005} & 0.0001 \\
 &  &  & 3 & 0.0265 & 0.0784 & 0.0056 & 0.0784 & 0.0047 & 0.0037 & 0.0007 & \textbf{0.0005} & 0.0001 \\
 &  & 0.8 & 1 & 0.0293 & 0.2390 & 0.0345 & 0.2394 & \textbf{0.0014} & 0.0061 & 0.0025 & 0.0026 & 0.0001 \\
 &  &  & 2 & 0.0384 & 0.0956 & 0.0287 & 0.0956 & 0.0039 & 0.0036 & 0.0017 & \textbf{0.0008} & 0.0001 \\
 &  &  & 3 & 0.0189 & 0.1310 & 0.0225 & 0.1310 & 0.0061 & 0.0068 & 0.0037 & \textbf{0.0013} & 0.0001 \\ \hline
AVar & 100 & 0.6 & 1 & 0.0189 & 0.0204 & \textbf{0.0033} & 1.0726 & 0.0042 & 0.0042 & 0.0048 & 0.0052 & 0.0051 \\
 &  &  & 2 & 0.0203 & 0.0230 & \textbf{0.0029} & 0.1082 & 0.1080 & 0.0043 & 0.0042 & 0.0050 & 0.0051 \\
 &  &  & 3 & 0.0194 & 0.0219 & \textbf{0.0028} & 0.0217 & 0.0044 & 0.0044 & 0.0047 & 0.0046 & 0.0051 \\
 &  & 0.8 & 1 & 0.0248 & 0.0397 & \textbf{0.0036} & 0.0388 & 0.0120 & 0.0039 & 0.0044 & 0.0050 & 0.0051 \\
 &  &  & 2 & 0.0268 & 0.0414 & \textbf{0.0018} & 0.0413 & 0.0073 & 0.0075 & 0.0038 & 0.0047 & 0.0051 \\
 &  &  & 3 & 0.0254 & 0.0371 & \textbf{0.0019} & 0.0371 & 0.0095 & 0.0078 & 0.0050 & 0.0043 & 0.0051 \\ \cmidrule{2-13}
 & 500 & 0.6 & 1 & 0.0034 & 0.0041 & \textbf{0.0007} & 0.0042 & 0.0008 & 0.0008 & 0.0009 & 0.0010 & 0.0011 \\
 &  &  & 2 & 0.0035 & 0.0045 & \textbf{0.0006} & 0.0045 & 0.0008 & 0.0008 & 0.0009 & 0.0009 & 0.0011 \\
 &  &  & 3 & 0.0037 & 0.0043 & \textbf{0.0006} & 0.0043 & 0.0009 & 0.0009 & 0.0010 & 0.0010 & 0.0011 \\
 &  & 0.8 & 1 & 0.0051 & 0.0072 & \textbf{0.0007} & 0.0070 & 0.0044 & \textbf{0.0007} & 0.0009 & 0.0010 & 0.0011 \\
 &  &  & 2 & 0.0049 & 0.0082 & \textbf{0.0003} & 0.0082 & 0.0013 & 0.0014 & 0.0008 & 0.0009 & 0.0011 \\
 &  &  & 3 & 0.0046 & 0.0076 & \textbf{0.0004} & 0.0076 & 0.0019 & 0.0016 & 0.0011 & 0.0009 & 0.0011 \\ \hline
MISE & 100 & 0.6 & 1 & 0.0489 & 0.1166 & 0.0122 & 1.1971 & 0.0076 & 0.0069 & \textbf{0.0057} & 0.0058 & 0.0052 \\
 &  &  & 2 & 0.0506 & 0.1098 & 0.0103 & 0.1965 & 0.1127 & 0.0074 & \textbf{0.0050} & 0.0055 & 0.0052 \\
 &  &  & 3 & 0.0466 & 0.1005 & 0.0084 & 0.1005 & 0.0091 & 0.0081 & 0.0055 & \textbf{0.0050} & 0.0052 \\
 &  & 0.8 & 1 & 0.0526 & 0.2776 & 0.0382 & 0.2768 & 0.0132 & 0.0100 & \textbf{0.0070} & 0.0074 & 0.0052 \\
 &  &  & 2 & 0.0638 & 0.1355 & 0.0303 & 0.1355 & 0.0110 & 0.0109 & \textbf{0.0053} & 0.0054 & 0.0052 \\
 &  &  & 3 & 0.0450 & 0.1692 & 0.0240 & 0.1692 & 0.0156 & 0.0146 & 0.0088 & \textbf{0.0055} & 0.0052 \\ \cmidrule{2-13}
 & 500 & 0.6 & 1 & 0.0340 & 0.0994 & 0.0094 & 0.0996 & 0.0045 & 0.0036 & 0.0018 & \textbf{0.0016} & 0.0012 \\
 &  &  & 2 & 0.0353 & 0.0917 & 0.0079 & 0.0917 & 0.0044 & 0.0040 & 0.0017 & \textbf{0.0014} & 0.0012 \\
 &  &  & 3 & 0.0302 & 0.0827 & 0.0062 & 0.0827 & 0.0056 & 0.0046 & 0.0016 & \textbf{0.0015} & 0.0012 \\
 &  & 0.8 & 1 & 0.0344 & 0.2462 & 0.0352 & 0.2464 & 0.0058 & 0.0069 & \textbf{0.0034} & 0.0035 & 0.0012 \\
 &  &  & 2 & 0.0433 & 0.1037 & 0.0291 & 0.1037 & 0.0052 & 0.0049 & 0.0025 & \textbf{0.0017} & 0.0012 \\
 &  &  & 3 & 0.0235 & 0.1385 & 0.0229 & 0.1385 & 0.0080 & 0.0084 & 0.0047 & \textbf{0.0022} & 0.0012 \\ \bottomrule
\end{tabular}}
\end{table}

\begin{table}
\centering
{
\caption{{
Estimation accuracy for $\beta(t,\tau)$ at $\tau=0.5$ in the heterogeneous setting $(\eta_\beta,\eta_\varepsilon)=(1,1)$ (quantile-varying coefficients and heteroscedastic errors), with sample sizes $n\in\{100,500\}$ and $L=100$ observed time points, based on 500 replications under piecewise-constant zero inflation schemes. 
Specifically, Case 1 (2 equal segments) $\pi_i(t)=\pi_0$ on $[0,0.5)$ and $1-\pi_0$ on $[0.5,1]$; Case 2 (5 equal segments) $\pi_i(t)=\pi_0$ on $[0,0.2)\cup[0.4,0.6)\cup[0.8,1]$ and $1-\pi_0$ on $[0.2,0.4)\cup[0.6,0.8)$; Case 3 (3 unequal segments) $\pi_i(t)=\pi_0$ on $[0,0.1)\cup[0.6,1]$ and $1-\pi_0$ on $[0.1,0.6)$.
Here $\widehat{M}$ denotes the working number of segments assumed by BE-ZIME during estimation. Bolded values denote the best performance (excluding the Oracle).
}}
\label{tab:s2beta1eta11}
\begin{tabular}{cccc|ccccccccc}
\toprule
& $n$ & $\pi_0$ & Case & Naive & Average & P-ZIPMM & BE-ME &
$\widehat{M}=1$ & $\widehat{M}=2$ & $\widehat{M}=5$ & $\widehat{M}=10$ &  Oracle \\ \midrule
ABias$^2$ & 100 & 0.6 & 1 & 0.0289 & 0.0959 & 0.0087 & 0.1400 & 0.0030 & 0.0023 & 0.0008 & \textbf{0.0006} & 0.0001 \\
 &  &  & 2 & 0.0298 & 0.0861 & 0.0073 & 0.0879 & 0.0032 & 0.0029 & 0.0007 & \textbf{0.0005} & 0.0001 \\
 &  &  & 3 & 0.0269 & 0.0788 & 0.0050 & 0.0793 & 0.0046 & 0.0036 & 0.0006 & \textbf{0.0003} & 0.0001 \\
 &  & 0.8 & 1 & 0.0281 & 0.2332 & 0.0344 & 0.2343 & \textbf{0.0009} & 0.0062 & 0.0024 & 0.0023 & 0.0001 \\
 &  &  & 2 & 0.0365 & 0.0929 & 0.0285 & 0.0929 & 0.0035 & 0.0033 & 0.0014 & \textbf{0.0007} & 0.0001 \\
 &  &  & 3 & 0.0188 & 0.1299 & 0.0215 & 0.1299 & 0.0066 & 0.0066 & 0.0038 & \textbf{0.0009} & 0.0001 \\ \cmidrule{2-13}
 & 500 & 0.6 & 1 & 0.0301 & 0.0967 & 0.0086 & 0.0966 & 0.0037 & 0.0028 & 0.0008 & \textbf{0.0006} & 0.0001 \\
 &  &  & 2 & 0.0309 & 0.0873 & 0.0071 & 0.0870 & 0.0035 & 0.0031 & 0.0007 & \textbf{0.0004} & 0.0001 \\
 &  &  & 3 & 0.0268 & 0.0791 & 0.0056 & 0.0791 & 0.0045 & 0.0034 & 0.0006 & \textbf{0.0004} & 0.0001 \\
 &  & 0.8 & 1 & 0.0292 & 0.2416 & 0.0342 & 0.2413 & \textbf{0.0013} & 0.0059 & 0.0024 & 0.0026 & 0.0001 \\
 &  &  & 2 & 0.0376 & 0.0957 & 0.0288 & 0.0957 & 0.0038 & 0.0035 & 0.0016 & \textbf{0.0008} & 0.0001 \\
 &  &  & 3 & 0.0188 & 0.1283 & 0.0225 & 0.1283 & 0.0065 & 0.0067 & 0.0037 & \textbf{0.0012} & 0.0001 \\ \hline
AVar & 100 & 0.6 & 1 & 0.0351 & 0.0658 & \textbf{0.0127} & 2.2765 & 0.0157 & 0.0155 & 0.0181 & 0.0199 & 0.0236 \\
 &  &  & 2 & 0.0358 & 0.0706 & \textbf{0.0112} & 0.2533 & 0.0155 & 0.0161 & 0.0164 & 0.0190 & 0.0236 \\
 &  &  & 3 & 0.0340 & 0.0703 & \textbf{0.0110} & 0.0694 & 0.0169 & 0.0165 & 0.0185 & 0.0193 & 0.0236 \\
 &  & 0.8 & 1 & 0.0450 & 0.1216 & \textbf{0.0108} & 0.1225 & 0.0370 & 0.0142 & 0.0158 & 0.0182 & 0.0236 \\
 &  &  & 2 & 0.0504 & 0.1365 & \textbf{0.0045} & 0.1364 & 0.0265 & 0.0269 & 0.0135 & 0.0167 & 0.0236 \\
 &  &  & 3 & 0.0516 & 0.1225 & \textbf{0.0062} & 0.1225 & 0.0351 & 0.0294 & 0.0196 & 0.0169 & 0.0236 \\ \cmidrule{2-13}
 & 500 & 0.6 & 1 & 0.0063 & 0.0138 & \textbf{0.0028} & 0.0139 & 0.0034 & 0.0032 & 0.0038 & 0.0043 & 0.0053 \\
 &  &  & 2 & 0.0065 & 0.0150 & \textbf{0.0022} & 0.0150 & 0.0033 & 0.0034 & 0.0037 & 0.0039 & 0.0053 \\
 &  &  & 3 & 0.0067 & 0.0144 & \textbf{0.0024} & 0.0142 & 0.0036 & 0.0038 & 0.0042 & 0.0042 & 0.0053 \\
 &  & 0.8 & 1 & 0.0095 & 0.0235 & \textbf{0.0021} & 0.0232 & 0.0091 & 0.0028 & 0.0034 & 0.0038 & 0.0053 \\
 &  &  & 2 & 0.0093 & 0.0275 & \textbf{0.0008} & 0.0275 & 0.0053 & 0.0056 & 0.0028 & 0.0034 & 0.0052 \\
 &  &  & 3 & 0.0095 & 0.0263 & \textbf{0.0013} & 0.0263 & 0.0077 & 0.0064 & 0.0043 & 0.0034 & 0.0053 \\ \hline
MISE & 100 & 0.6 & 1 & 0.0640 & 0.1617 & 0.0214 & 2.4165 & 0.0186 & \textbf{0.0179} & 0.0189 & 0.0205 & 0.0238 \\
 &  &  & 2 & 0.0656 & 0.1567 & 0.0185 & 0.3413 & 0.0186 & 0.0190 & \textbf{0.0172} & 0.0195 & 0.0238 \\
 &  &  & 3 & 0.0609 & 0.1491 & \textbf{0.0160} & 0.1488 & 0.0216 & 0.0201 & 0.0191 & 0.0196 & 0.0238 \\
 &  & 0.8 & 1 & 0.0730 & 0.3548 & 0.0453 & 0.3568 & 0.0379 & 0.0204 & \textbf{0.0182} & 0.0205 & 0.0238 \\
 &  &  & 2 & 0.0869 & 0.2294 & 0.0330 & 0.2293 & 0.0299 & 0.0302 & \textbf{0.0149} & 0.0174 & 0.0238 \\
 &  &  & 3 & 0.0704 & 0.2524 & 0.0277 & 0.2524 & 0.0417 & 0.0361 & 0.0235 & \textbf{0.0178} & 0.0238 \\ \cmidrule{2-13}
 & 500 & 0.6 & 1 & 0.0364 & 0.1104 & 0.0114 & 0.1105 & 0.0071 & 0.0060 & \textbf{0.0046} & 0.0049 & 0.0054 \\
 &  &  & 2 & 0.0374 & 0.1023 & 0.0093 & 0.1020 & 0.0069 & 0.0066 & \textbf{0.0044} & \textbf{0.0044} & 0.0054 \\
 &  &  & 3 & 0.0335 & 0.0934 & 0.0080 & 0.0934 & 0.0080 & 0.0071 & 0.0048 & \textbf{0.0046} & 0.0054 \\
 &  & 0.8 & 1 & 0.0388 & 0.2651 & 0.0363 & 0.2644 & 0.0105 & 0.0087 & \textbf{0.0058} & 0.0064 & 0.0053 \\
 &  &  & 2 & 0.0468 & 0.1232 & 0.0296 & 0.1232 & 0.0092 & 0.0091 & 0.0044 & \textbf{0.0042} & 0.0053 \\
 &  &  & 3 & 0.0282 & 0.1546 & 0.0238 & 0.1546 & 0.0142 & 0.0131 & 0.0080 & \textbf{0.0047} & 0.0053 \\ \bottomrule
\end{tabular}}
\end{table}

\begin{table}
    \centering
    \caption{Descriptive statistics for the childhood obesity study participants after data cleaning. 
    ``Mean (sd)'' is shown for the continuous variable (Age), and ``Count (\%)'' indicates the number (and percentage) of participants with a value of 1 for each binary variable.}
    \label{tab:real_cov}
    \begin{tabular}{lcl}
    \toprule
        Variable & Mean(sd)/Count(\%) & Description \\ 
        \midrule
        Age & 7.97 (0.754) & Age in years.\\
        Ethnicity & 135 (78.035) & Indicator for White (1) vs.\ Other (0).\\
        Gender & 88 (50.867) & Indicator for Girl (1) vs.\ Boy (0).\\
        School$_2$ & 97 (56.069) & Indicator for attending School~2 (1).\\
        School$_3$ & 48 (27.746) & Indicator for attending School~3 (1).\\
        Treatment & 66 (38.150) & Indicator for stand-biased desks (1).\\
    \bottomrule
\end{tabular}
\end{table}

\begin{table}
    \centering
    \caption{\(p\)-values from a global wild-bootstrap-based test of the null hypothesis \(H_0\colon \beta(t, \tau = 0.5) \equiv 0\) for each method. Bolded values indicate statistical significance at the 5\% level. ``EE-Method'' and ``SC-Method'' refer to estimates based on corrected energy expenditure and step counts, respectively.}
    \label{tab:real_pv}
    \begin{tabular}{lr}
    \toprule
    Method & p-value \\
    \midrule
    SC-Naive       & \textbf{0.016} \\
    SC-Average     & 0.513 \\
    SC-P-ZIPMM     & \textbf{0.039} \\
    SC-BE-ZIME-3   & \textbf{0.009} \\
    EE-P-LMM       & \textbf{0.015} \\
    \bottomrule
    \end{tabular}
\end{table}
}

\end{document}